\documentclass{emulateapj}

\newcommand{\ltsima}{$\; \buildrel < \over \sim \;$}
\newcommand{\lsim}{\lower.5ex\hbox{\ltsima}}
\newcommand{\gtsima}{$\; \buildrel > \over \sim \;$}
\newcommand{\gsim}{\lower.5ex\hbox{\gtsima}}

\shorttitle{Morphological quenching of star formation}
\shortauthors{Martig et al.}

\begin{document}

\title{Morphological quenching of star formation: \\
making early-type galaxies red}

\author{Marie Martig, Fr\'{e}d\'{e}ric Bournaud, Romain Teyssier,} 
\affil{CEA, IRFU, SAp. F-91191 Gif-sur-Yvette, France.}
\affil{Laboratoire AIM, CNRS, CEA/DSM, Universit{\'e} Paris Diderot, France.}
\email{marie.martig@cea.fr}
\and
\author{Avishai Dekel}
\affil{Racah Institute of Physics, The Hebrew University, Jerusalem 91904, Israel.}

\begin{abstract}
We point out a natural mechanism for quenching of star formation in early-type
galaxies. It automatically links the color of a galaxy with its morphology
and does not require gas consumption, removal or termination of gas supply.
Given that star formation takes place in gravitationally unstable gas disks,
it can be quenched when a disk becomes stable against fragmentation to bound
clumps. This can result from the growth of a stellar spheroid, for instance by mergers. We present the concept of morphological quenching (MQ) using standard disk instability analysis, and demonstrate its natural occurrence in a cosmological simulation using an efficient zoom-in technique. We show that the transition from a stellar disk to a spheroid can be sufficient to stabilize the gas disk, quench star formation, and turn an early-type galaxy red and dead while gas accretion continues. The turbulence necessary for disk stability can be stirred up by sheared perturbations within the disk in the absence of bound star-forming clumps. 
While other quenching mechanisms, such as gas stripping, AGN feedback, virial shock heating, and gravitational heating are limited to massive halos, the MQ can explain the appearance of red early-type galaxies also in halos less massive than $\sim 10^{12} {\rm M}_\odot$.
The dense gas disks observed in some of today's red ellipticals may be the relics of this mechanism, whereas red galaxies with quenched gas disks could be more frequent at high redshift.
\end{abstract}


\section{Introduction}
In the low redshift Universe, the variety of morphological types of galaxies is accompanied by a color bimodality \citep{Roberts1994}. Spiral galaxies have low bulge-to-disk ratios and are actively star forming, with blue colors. Early-type galaxies (ETGs), which are spheroid-dominated lenticulars and ellipticals, are red, non- or almost non-star-forming. The formation of disk galaxies could be explained by a variety of processes, from smooth gas accretion and internal evolution \citep{E05, BEE07, G08, D09, DSC09} to the early merging of gas-rich galaxies \citep{robertson05, SH05, governato06}. There is a larger consensus that the transformation of spiral disks into elliptical galaxies is largely driven by major and minor galaxy mergers \citep{BH92,NB03,B04,B07}. The morphological transformation of spirals into ellipticals thus has to be accompanied by a ``quenching'' of their star formation, which shifts them to the Red Sequence.

This quenching of star formation can be obtained by consuming the cold gas reservoir. However, simulations suggest that this might not be the case for typical mergers \citep{diMatteo2007,diMatteo2008}. In standard pictures, some extra quenching processes are then needed to completely remove the cold gas reservoir of these post-merger galaxies and to shut down any fresh supply of cold gas. Such processes include (i) a cluster environment in which the cold gas is stripped and no cold gas can be found in the neighborhood \citep[but see][]{tonessen, salome, revaz}, (ii) heating by virial shocks of gas infalling in a massive dark matter halo \citep{BD03,DB06,Birnboim2007,cattaneo2008,Ocvirk2008,Keres2009}, (iii) energy feedback from an Active Galactic Nucleus (AGN), which could itself be triggered by galaxy mergers \citep{Hopkins2005,dimatteo,cattaneo1, cattaneo2, cattaneo2009, Johansson2009a, Schawinski2009} (iv) and gravitational heating of a gas disk by clumpy mass infall \citep{DB08} and minor mergers \citep{Johansson2009b}. While AGN feedback could participate in the reddening of massive galaxies and nuclear activity is frequently observed at high redshift \citep{Nesvadba2008}, there is a lack of direct evidence that this is the looked-for quenching mechanism -- AGN jets could be too collimated to affect most cold gas disks, and AGNs could theoretically trigger the star formation as well. Gas heating by AGN feedback and virial shock is furthermore limited to halos above $\sim$10$^{12}$ M$_\odot$ \citep{DB08}.

The mechanisms listed above therefore struggle to explain the reddening of field ETGs in halos below this critical virial mass \citep[e.g.,][]{weinmann1, weinmann2, hopkins}.
In this paper, we propose a new mechanism called ``morphological quenching'' (MQ), in which star formation is quenched even for ETGs in halos of moderate masses. Contrary to the mechanisms proposed so far, we show that an ETG can spontaneously turn red without having its cold gas removed, and could even contain amounts of cold gas similar to those of some spiral galaxies. This gas would be stabilized against star formation by the morphological transition from a rotating stellar disk to a pressure-dominated spheroid, which induces a steeper potential well and reduces the disk self-gravity. Observations of disk galaxies with various bulge fractions \citep{kennicutt89} had already shown that bulge dominated galaxies form stars less efficiently than disk-dominated ones. We propose that this morphological quenching of star formation could participate in the reddening of elliptical and lenticular galaxies.

This mechanism is found to naturally occur in numerical simulations of galaxy evolution in the $\Lambda$-CDM context. We use a new technique for cosmological zoom-in simulations, in which the cosmological history of a given galaxy (merger tree and smooth accretion) is recorded, and re-simulated at high-resolution. Our simulations suggest that weak gravitational instabilities in the gas disk (spiral arms) can trigger the turbulence necessary for stability while not allowing dense bound clouds to form, preventing efficient star formation.

Morphological quenching is in agreement with observations showing that some Local red ETGs host significant amounts of atomic or molecular gas with high densities, sometimes in amounts similar to those of the Milky-Way gas disk, but with only very inefficient to no star formation at all and red integrated colors \citep{morganti06,crocker08b,youngbl08,donovan09}. This process could be even more important at high redshift where a red sequence is likely already in place \citep{Bell2004, Cirasuolo2007} while gas fractions in early-type galaxies are expected to be higher since gas infall rates are high and most clusters are not virialized yet.

Our simulations also indicate that a reddened elliptical galaxy could turn back again into a blue star-forming disk galaxy provided that it re-accretes sufficient amounts of cold gas. This transition back to the blue sequence requires high rates of cold gas accretion, probably making this process infrequent at $z<1$.

The MQ mechanism is described in Section~2. Cosmological models illustrating the formation of ETGs with morphological quenching are presented in Section~3, and the stability of their gas disk against star formation is studied in Section~4. Implications for various observations are discussed in Section~5.

\section{Morphological quenching}

Gravitational instability is a necessary condition for efficient star formation
in a galactic disk \citep[e.g.,][]{Elmegreen2002,Li2005,Kawata2007}
The instability of a single-component, rotating disk at radius
$r$ is characterized by the Toomre $Q$ parameter \citep{Toomre1964}:

\begin{equation}\label{defQ}
 Q=\frac{\kappa \sigma}{\epsilon G \Sigma} \, .
\end{equation}
Here $\sigma$ is the one-dimensional velocity dispersion and $\Sigma$ is
the surface density of the disk. $\kappa$ is the epicyclic frequency, which is linked to the depth of the gravitational potential, and is of the order of the angular velocity $\Omega$.
The factor $\epsilon$ is 3.36 for a thin stellar disk and $\pi$ for a gas disk.
Axisymmetric instabilities, namely rings that break into clumps, can grow in
the disk if $Q<1$.

The instability of a two-component (gas and stars) disk is characterized by
an effective Toomre parameter \citep{jog-2comp,elmegreen-2comp} defined by
\begin{equation}\label{stab}
Q^{-1} = \alpha_g Q_{\rm g}^{-1}+ \alpha_s Q_{\rm s}^{-1} \, ,
\end{equation}
where the $Q$ parameters for the gas (g) and for the stars (s) are computed
separately for each component using Equation~\ref{defQ}. The coefficients are
\begin{equation}
\alpha_{\rm g}= \frac{2\sigma_{\rm gs}q}{1+\sigma_{\rm gs}^2q^2}
\end{equation}
and
\begin{equation}
 \alpha_{\rm s}= \frac{2q}{1+q^2} \, ,
\end{equation}
where $\sigma_{\rm gs}\equiv \sigma_{\rm g}/ \sigma_{\rm s}$ and the
dimensionless wave number $q \equiv k \sigma_{\rm s}/\kappa $ is the one
minimizing $Q$. These coefficients are of order unity in spiral galaxies.

The stellar surface density $\Sigma_{\rm s}$ in Equation~\ref{defQ} refers
to the stars that co-rotate in a relatively thin disk and not to the stars
that populate the spheroids, bulge or halo. In particular, for an elliptical
galaxy with all the stars in a pressure-supported spheroid,
$Q_{\rm s} \sim \infty$ . A lenticular galaxy with a thick, kinematically hot
stellar disk embedded in a massive bulge has $Q_{\rm s} >> 1$.

The stability of a disk for clump-forming instabilities and subsequent star
formation has indeed been shown to be properly characterized by the effective
$Q$ parameter both in observations (Wong \& Blitz 2002, Leroy et al. 2008)
and in simulations (Li et al. 2005). A disk with a sufficiently low gas
surface density and/or a sufficiently high epicyclic frequency has
$Q_{\rm g}>1$, which, by itself, would have indicated stability for
axisymmetric modes. However, for a given $\Sigma_{\rm g}$ and $\kappa$,
the presence of a stellar disk component enhances the self-gravity within
the perturbation and can destabilize the disk, pulling the effective $Q$
parameter to below unity.
For example, typical low-redshift disk galaxies have $\sigma_{\rm s} >> \sigma_{\rm g}$
as well as $\Sigma_{\rm s} >> \Sigma_{\rm g}$, so $Q_{\rm s}$ could be comparable
to $Q_{\rm g}$ and both can thus play a similar role in determining the
disk instability. The same likely holds at high redshift, where
$\Sigma_{\rm g} \simeq \Sigma_{\rm s}$ but
$\sigma_{\rm g} \simeq \sigma_{\rm s}$ \citep[see for instance][]{FS09}.
The $Q<1$ condition ensure the formation of gas clumps in which new stars form, without implying that the pre-existing stars collapse in these gas clumps.

Furthermore, a necessary requirement for star formation is
that the internal gravity in gas clouds overcomes the disruptive
galactic tidal field. This requires a three-dimensional disk density above
the critical tidal density \citep{hunter}
\begin{equation}
\rho_{\mathrm{tidal}} = - \frac{3 \Omega r}{2 \pi G}
\frac{\mathrm{d} \Omega}{\mathrm{d} r} \, .
\end{equation}
Note that the relevant disk density involves both the gaseous and stellar components
of the disk; the disk stars may help the self-gravity of the clumps
overcome the external tidal forces even in cases where the disk gas is not dense enough for the purpose.

\begin{figure*}
\centering
\includegraphics[width=8cm]{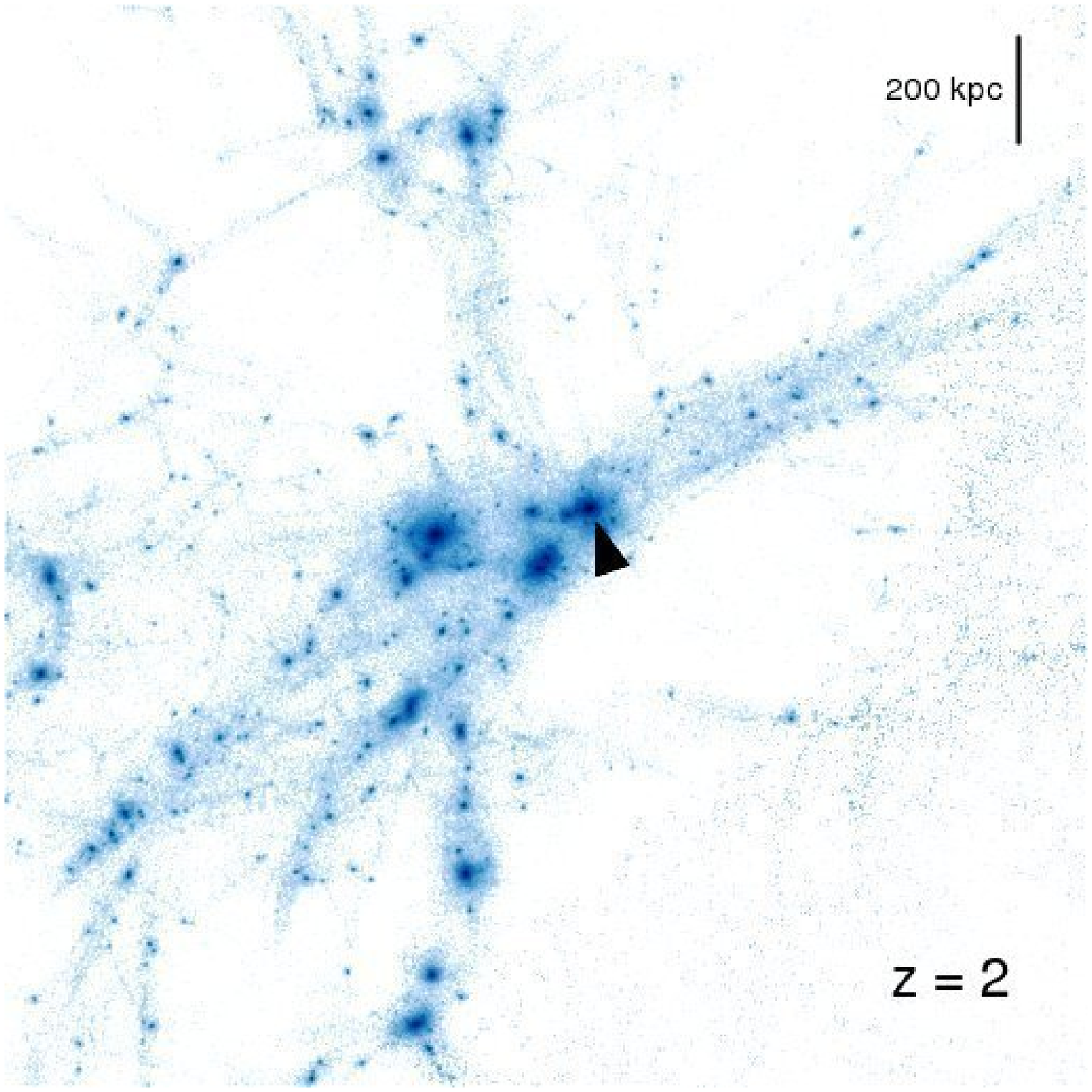}
\includegraphics[width=8cm]{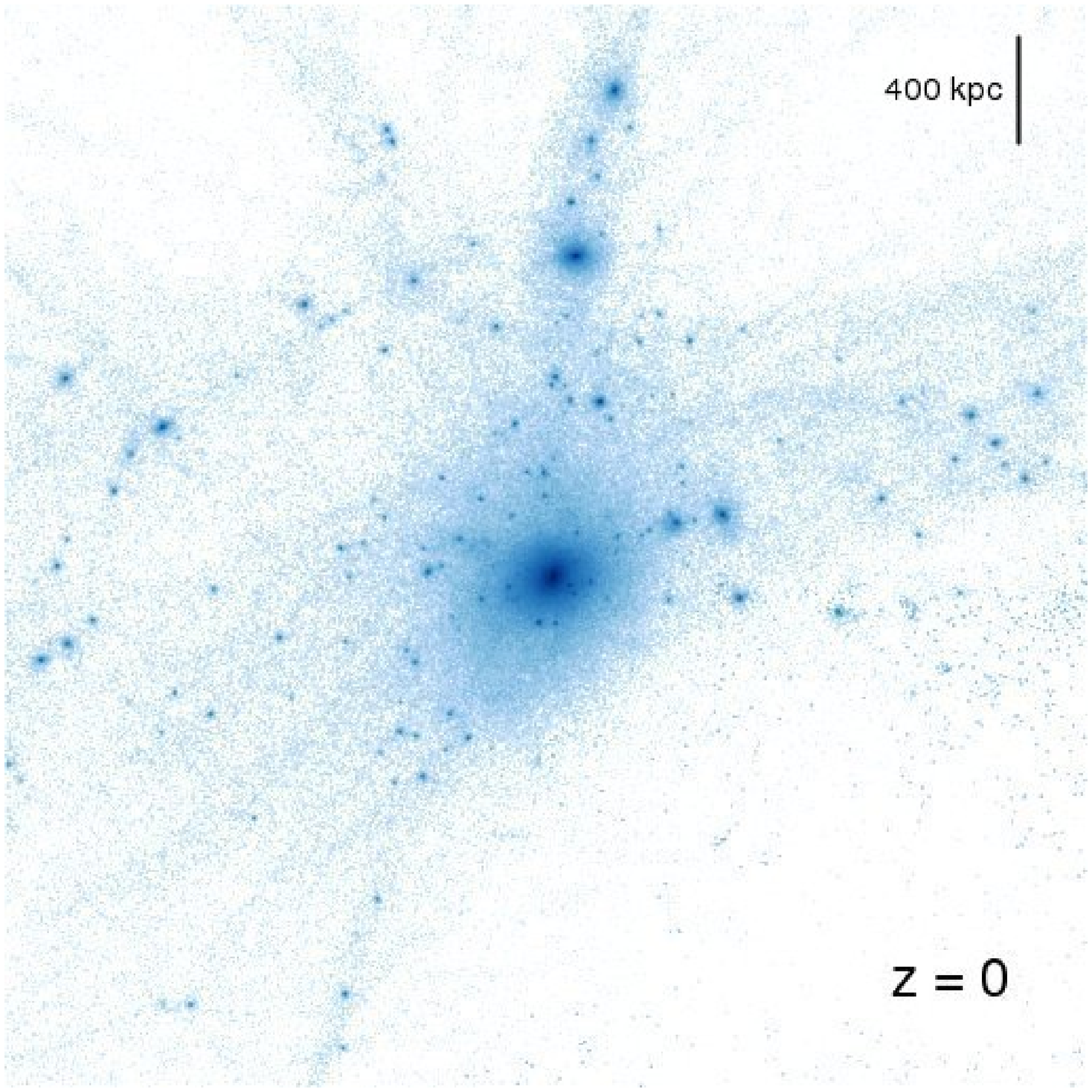}
\caption{Large-scale dark-matter distribution in the cosmological simulation.
Left: at $z=2$ in a box of side 2~Mpc.
Right: at $z=0$ in a box of side 4~Mpc.
The halo on which we zoom in is marked by an arrow in the left panel and
is at the center in the right panel.
The $z=2$ snapshot shows neighboring halos that will merge with our halo
at $z\sim 1.5$. The $z=0$ snapshot follows a major merger that occurred at 
$z=0.2$. The halo is continuously fed by gas streams that follow the
dark-matter filaments seen in these maps.
}\label{FigCosmo}
\end{figure*}

The basic idea of morphological quenching is that the star formation
in a gaseous disk could be severely suppressed even if the disk is relatively massive once it is embedded in an early-type galaxy, that is in an elliptical or a bulge-dominated lenticular.
A similar gas disk could be effectively star-forming in a spiral galaxy
where most of the stellar mass is in a rotating disk.
As a result, the former type of galaxies would naturally have red colors, while the latter would be
bluer. The difference in disk stability arises from two main effects:
(i) the higher concentration of the stellar mass in an ETG, affecting $Q$ via $\kappa$, and (ii) the contribution of the stellar disk to the
self-gravity in a spiral, affecting $Q$ via $\Sigma_{\rm s}$.
From another perspective, when the stellar disk becomes a spheroid in an ETG,
the self-gravity of the gas disk that is left behind can fail to balance
the disruptive tidal forces, which prevents the assembly of star-forming
clumps.

Bulge-dominated galaxies are therefore likely to form stars less efficiently
than disk-dominated ones, as observed by \cite{kennicutt89} for disk
galaxies with various bulge fractions.
We here propose that this effect could actually lead to quenching
of star formation in E and S0 galaxies and make them red.
 Furthermore, in the following, we show in numerical simulations that morphological quenching can naturally occur in the cosmological context, and that gas
disks with masses up to several $10^9$~M$_{\sun}$ can be stabilized in ETGs.
Our simulations suggest that a rather high turbulent speed ($\sigma_{\rm g}$) can be
maintained in a stabilized gas disk in an ETG (see Section~4), which provides
the pressure support necessary for stability.

\section{Numerical simulations}

\subsection{Method}

To study galaxy evolution in the cosmological context, the standard technique consists in running full cosmological simulations with spatial and mass resolutions increased in a certain ``zoom-in'' area \citep{Katz1993,Governato2004,SC05,Naab2007,Joung2009}. A constraint is that all particles ending-up in the target galaxy should be low-mass, high-resolution ones. This makes uneasy, if possible at all, to simulate galaxies with high accretion and merger rates at all redshifts, while more isolated galaxies or massive galaxies with early assemblies are more easily modeled this way.

We here use an original approach where we extract merger and accretion histories (and geometry) for a given halo/galaxy in a $\Lambda$-CDM cosmological simulation (see Appendix A.1), and then re-simulate theses histories at much higher resolution, replacing each halo by a realistic ga containing gas, stars and dark matter (Appendix A.2). The re-simulation is performed with a particle-mesh code where gas dynamics is modeled with a sticky particle scheme (Appendices A.2 and B).
\begin{figure*}
\centering
\includegraphics[width=17cm]{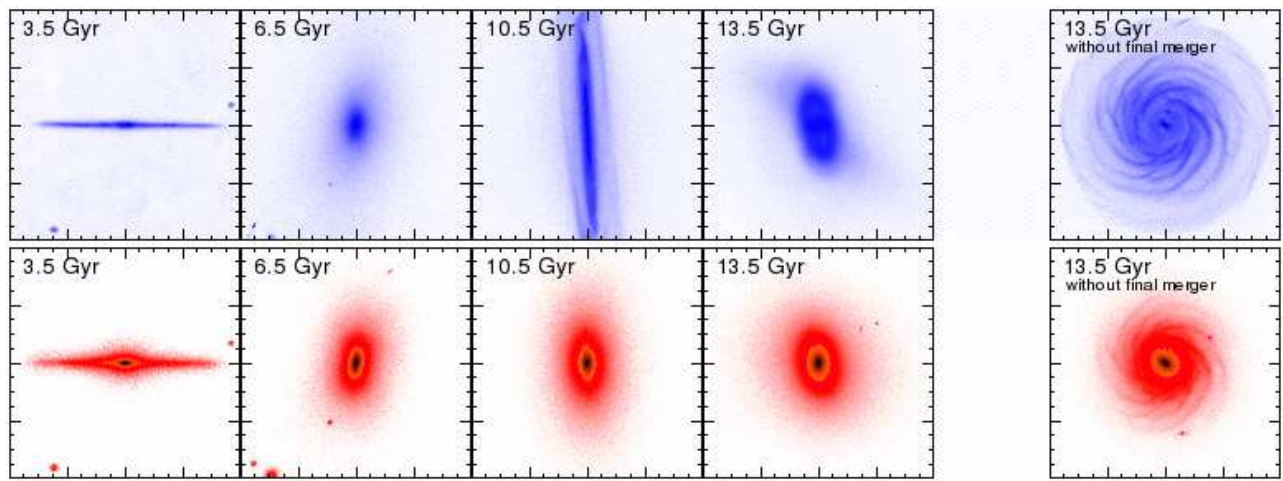}
\caption{
Evolution of the distribution of gas (top panels) and stars (bottom 
panels). The snapshot at $t=13.5$~Gyr is shown both for the original 
simulation and for its variant performed without the final major merger. 
The box size is 40~kpc in physical units.
The sequence of snapshots highlights the three phases of evolution.
Until $t=6.5$~Gyr, there is a gradual build-up of a stellar spheroid
due to a suite of mergers.
At $t=10.5$~Gyr, the system is toward the end of the MQ phase, showing
a large gas disk and a massive stellar spheroid.
At $t=13.5$~Gyr, after the major merger, the galaxy is an elliptical,
but if this merger is eliminated, it becomes an ETG with massive disk and 
spheroid components.
}\label{Evol}
\end{figure*}
This method takes into account not only every merger prescribed by the cosmological simulation, but also the smooth gas and dark matter infall along cold flows or cosmic filaments. A similar approach has been independently proposed by Kazantzidis et al. (2008), Read et al. (2008) and Villalobos \& Helmi (2008), but these authors account only for the main mergers and not for the diffuse accretion, and only performed collisionless simulations without any gas component, and sometimes even no baryonic component in satellite galaxies. Here we model both the main studied galaxy and all companions with dark matter, stars, and gas particles, and include all the accreted mass, from major and minor mergers to smooth accretion (Appendices A.2.1 and A.2.2). This allows us to study galaxy evolution in a realistic cosmological context from a high initial redshift ($z= 2$ in the present paper) down to $z = 0$ and account for the internal processes at high resolution: the maximal spatial resolution (and softening length) is 130~pc, while the mass resolution  is 1.4$\times$10$^5$~M$_{\sun}$ for stellar particles initially present in the galaxies, 2.1$\times$10$^4$~M$_{\sun}$ for gas particles and stellar particles formed during the simulation, and 4.4$\times$10$^5$ M$_{\sun}$ for dark matter particles.
The technique is described in detail in Appendix~A. Its main advantage is a reasonable computer cost for all kind of galaxies, even in dense environments and with high infall rates at all redshifts. Large samples of simulations with this technique will be studied in future papers.

\medskip

In the present paper, we focus our analysis on the color evolution of a galaxy similar to the Milky Way in terms of mass at $z=0$ and which does not belong to a rich group or a cluster of galaxies. The host halo is shown with its cosmological environment at $z=2$ and $z=$~0 on Figure~\ref{FigCosmo}. During the simulation, the halo mass grows from 2$\times 10^{11}$~M$_{\sun}$ at $z=2$ to 1.4$\times 10^{12}$~M$_{\sun}$ at $z=0$. This galaxy undergoes a first series of several minor mergers soon after $z=2$, four of which have mass ratios between 4:1 and 10:1. Then comes a quieter phase that is dominated by smooth dark matter and gas accretion, with an average gas accretion rate of 9~M$_{\sun}$ yr$^{-1}$, lasting from $z \simeq 1.2$ down to $z \simeq 0.2$. Finally, a major merger takes place at $z=0.2$ with a mass ratio of 1.5:1.
It is shortly preceded by a peak in the diffuse gas accretion rate (up to 70~M$_{\sun}$ yr$^{-1}$).

A case without the final major merger but the same history for other mergers and diffuse mass accretion is also studied. In this modified simulation, the whole mass corresponding to the incoming galaxy (both dark matter and baryons) is removed.

\subsection{Results}

\begin{figure*}
\centering
\includegraphics[width=8cm]{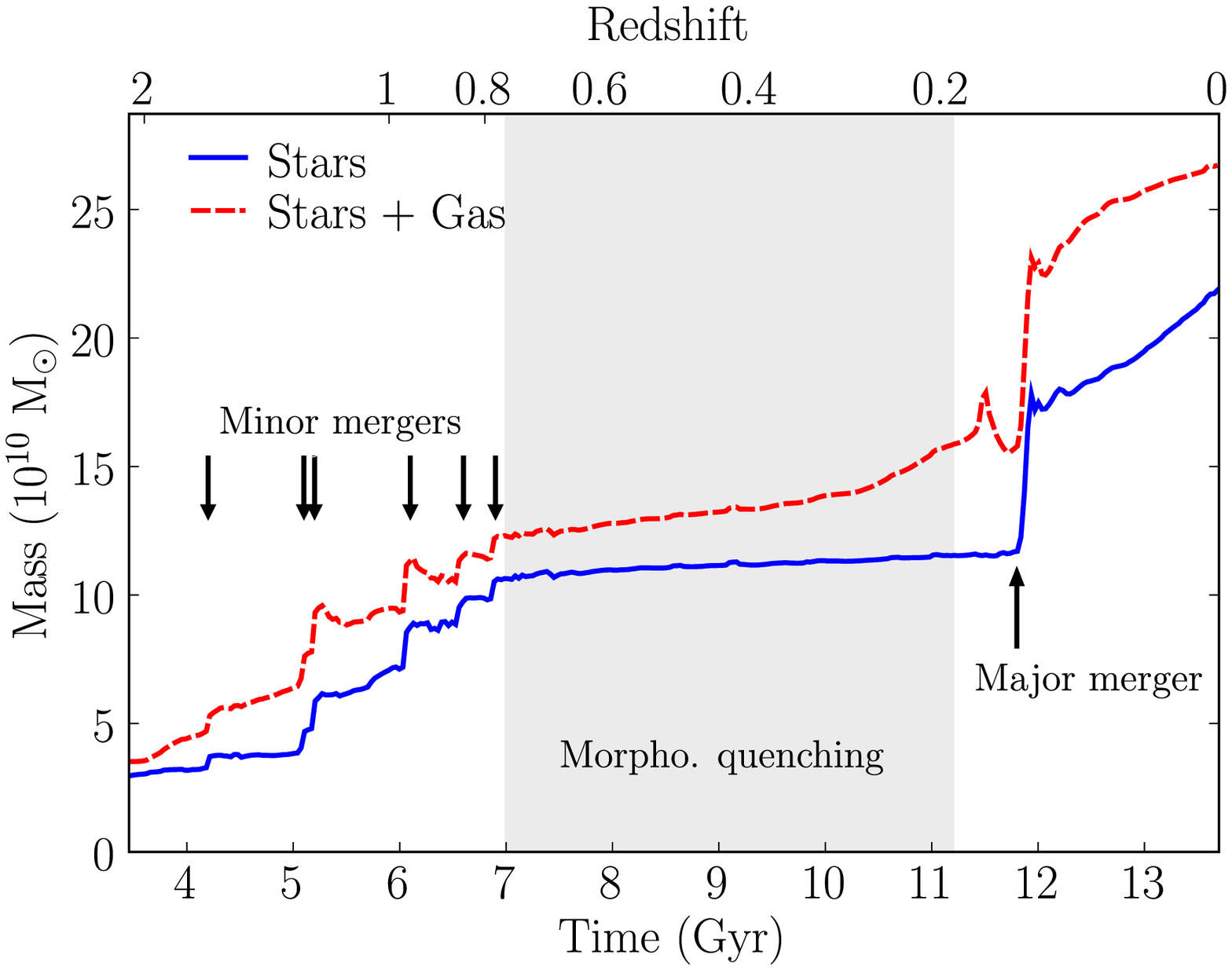}
\includegraphics[width=8cm]{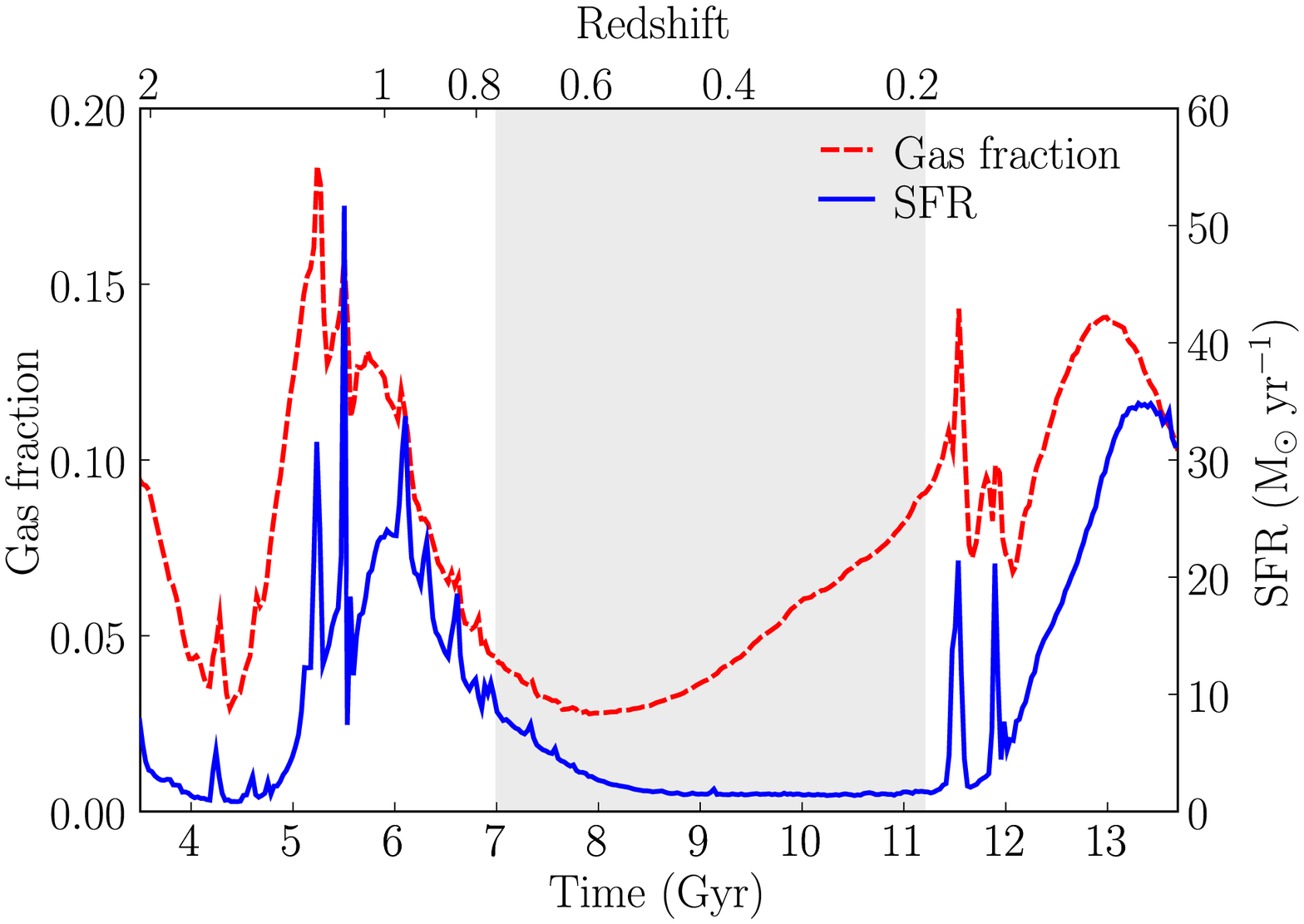}
\caption{Evolution of the stellar and gas components in the central
sphere of radius 25~kpc.
Left: mass in stars (solid, blue) and in baryons (dashed, red).
Right: the fraction of gas with density above $0.01$ M$_{\odot}$pc$^{-3}$
with respect to the total baryonic mass and the star formation rate.
The shaded area highlights the morphological-quenching phase,
where the SFR is suppressed and the stellar content is almost constant
while the gas content is increasing gradually due to continuous accretion.
The earlier and later phases are characterized by high gas fraction and high
SFR. The growth in the early phase is dominated by a sequence of minor mergers,
and in the late phase by a major merger.\vskip 0.5cm
}\label{mstar}
\end{figure*}

The morphological evolution from redshift~2 to redshift~0 is shown on Figure~\ref{Evol}. The evolution of stellar and gaseous masses, star formation rate (SFR) and integrated $B-V$ color are shown on Figures~\ref{mstar} and \ref{color}. These quantities are integrated in a sphere of radius 25 kpc around the main galaxy (see Appendix A.3 for details on the analysis technique).

\begin{figure}
\centering
\includegraphics[width=8.5cm]{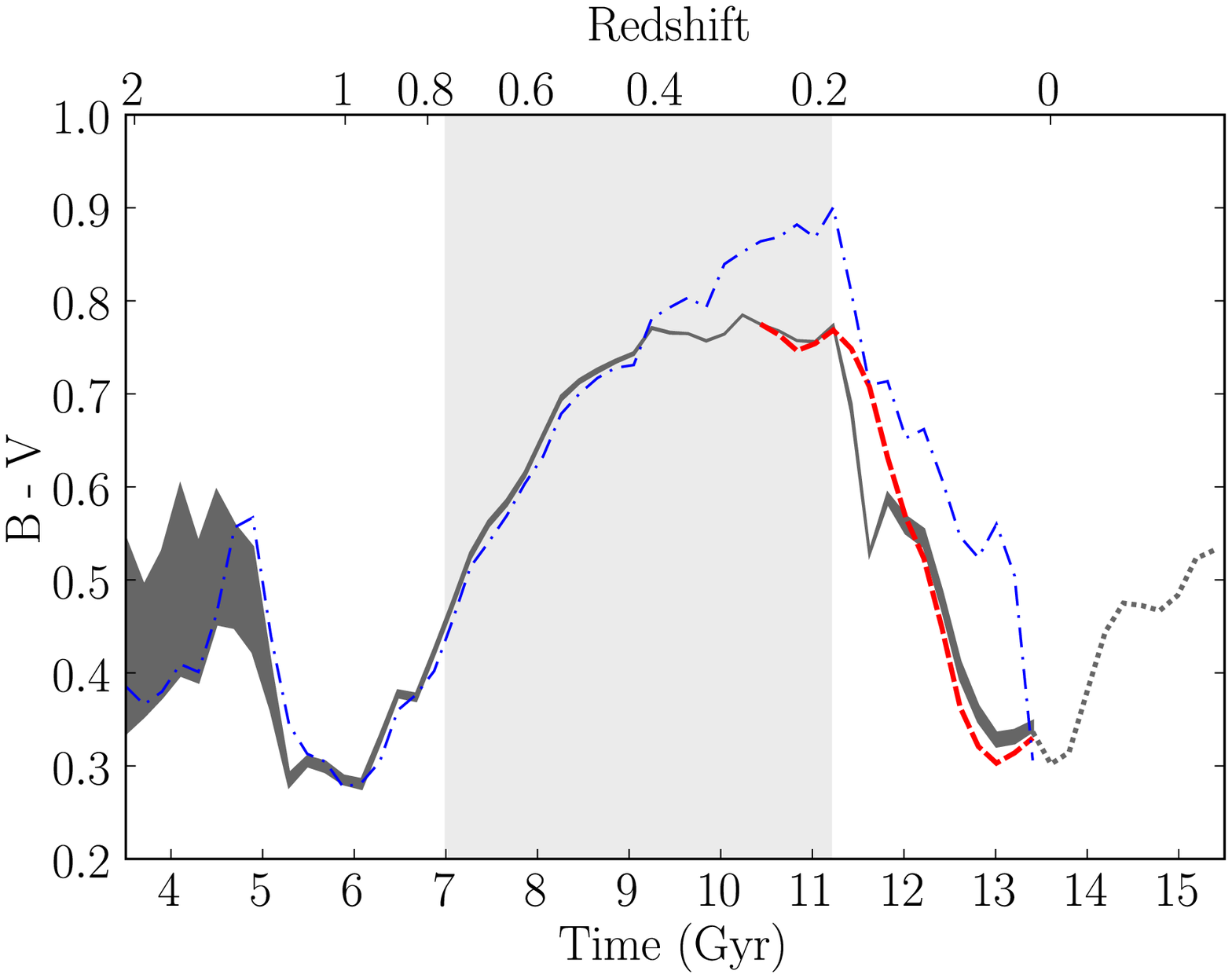}
\caption{Evolution of the $B-V$ color, showing the three phases as in Figure~3. 
The solid (dark grey) curve corresponds to the fiducial simulation.
The galaxy is blue during the early and late phases, and is reddened by half
a magnitude during the MQ phase.
The dark shaded area along the curve represents the range of colors 
obtained for different star formation histories (Appendix \ref{appendix_sim}). 
The dash-dotted (blue) line corresponds to a star
formation threshold at 0.25~M$_{\odot}$pc$^{-3}$ (Section~4.3),
showing a somewhat stronger quenching toward the end of the MQ phase.
The dashed line (red) refers to the case where the final major merger
is eliminated, showing only little differences. 
The dotted curve shows the evolution after $z=0$, where 
the system is about to enter a new quenched phase. 
The reddening associated with MQ is a robust phenomenon.
}\label{color}
\end{figure}
\subsubsection{Early mergers and spheroid buildup}

The simulation starts with a gas-rich disk galaxy (see Appendix A.2.1), forming stars actively in giant clumps, consistently with observations of high-redshift galaxies \citep{elmegreen07,G08}. A series of minor mergers takes place between redshifts~2 and~1, four of which have mass ratios between 4:1 and 10:1. These minor mergers transform the initial disk galaxy into an elliptical-like one, i.e., a stellar spheroid with a low rotational support ($V/\sigma<$1) \citep{B07}. The most important mergers occur at $z\sim 1.5$, and by $z\simeq 1$ the galaxy has an elliptical shape, even if still interacting with a few lower-mass companions.

This morphological transformation goes along with an episode of active star formation, reaching 50 M$_{\sun}$~yr$^{-1}$ (right panel of Figure~\ref{mstar}). At the same time, external gas is infalling along a filament, which maintains a high gas fraction (the right panel of Figure~\ref{mstar}) and sustains a high star formation rate during 2-3 Gyr.

As for the integrated colors, their evolution during the first 1.5~Gyr is sensitive to the assumptions made on the star formation history of each galaxy before its entry into the virial radius (see Appendix~A), which can be seen as an initial scatter in Figure~\ref{color}. Under most realistic assumptions, the initial galaxy is blue ($B-V<0.5$). The starbursts taking place during the minor mergers rapidly wash out any initial difference, and result in a very blue galaxy with $B-V=0.28$ at $t=6$~Gyr.

\subsubsection{Morphological quenching phase}

After the initial mergers, a much quieter period lasts from $z \simeq 1$ to $z=0.2$, during which no merger takes place other than with tiny satellites. This phase is quiescent in terms of merger events; however, diffuse gas is continuously infalling at a steady rate; the average gas accretion rate across the virial radius of the studied galaxy during this phase is 9~M$_{\sun}$~yr$^{-1}$.

The infalling gas adds to the gas mass left over after the initial mergers, so that the cold and dense ($\rho > 0.01$ M$_{\odot}$pc$^{-3}$) gas mass in the studied galaxy varies from 5.1$\times$10$^{9}$~M$_{\sun}$ at $t=8$~Gyr to $1.3 \times 10^{10}$~M$_{\sun}$ at $t=11$~Gyr (the left panel of Figure ~\ref{mstar}). During the quiescent phase, the galaxy thus remains relatively gas-rich, with a dense gas fraction increasing from 4\% to 8\% (with respect to the total baryonic mass inside a 25~kpc radius, see Figure~\ref{mstar}).

\begin{figure}
\centering
\includegraphics[width=8cm]{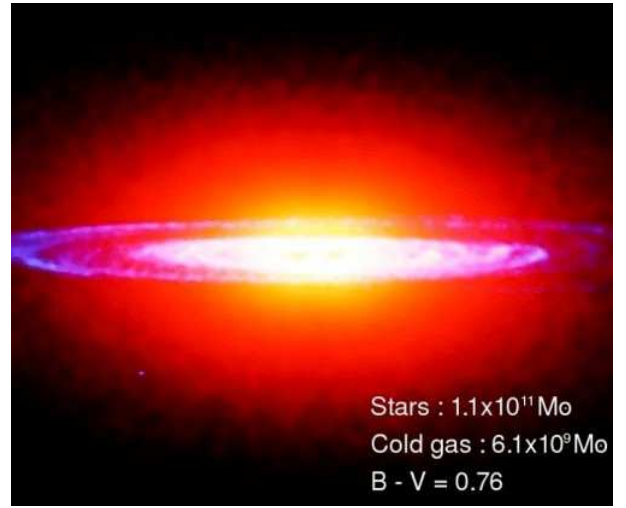}
\caption{
The galaxy during the morphological quenching phase at $t=10$~Gyr.
The box side is 30~kpc. The stellar mass distribution is
represented in yellow-red colors and the gas distribution in white-purple colors. 
The gas disk undergoing morphological quenching of star formation can be 
seen inside the spheroidal stellar component.
}\label{galax_10Gyr}
\end{figure}

This gas settles into a disk that slowly grows with time and acquires a tilt of almost 90 degrees with respect to the initial disk plane (Figure~\ref{Evol} and~\ref{galax_10Gyr}) because the main filament providing the infalling gas is almost perpendicular to the initial disk plane. This gas disk grows from the inside out: the density of the inner regions ($r < 2.5$~kpc) does not evolve much during the quiescent phase while the accreting gas settles in the outskirts of the disk .

\medskip

During this period, star formation is not completely suppressed but the star formation rate remains nearly constant at around 1.5~M$_{\sun}$~yr$^{-1}$, which is a low level considering the gas content of the galaxy, as shown in Figure \ref{mstar}: between $z=0.6$ and $z=0.2$ the gas fraction increases by a factor of 2 while the SFR remains low. For such gas fractions, and more importantly such gas surface densities (see also Section 4.1 and Table \ref{table_disks}), star formation rates above 10~M$_{\sun}$~yr$^{-1}$ could have been expected: for instance the galaxy at the beginning of the MQ phase has the same gas mass and gas surface density as the disk galaxy at $t=3.5$~Gyr but the latter forms stars at a rate of 9~M$_{\sun}$~yr$^{-1}$ (this will be discussed in more details in Section 4.1).

The low SFR during the MQ phase allows the formation a low-mass young stellar disk that never exceeds a few percent of the total stellar mass: on average 2\%, and at most 6\% at $t$=11~Gyr which is the end of the quiescent phase. Thus, most of the stars of this galaxy still belong to the spheroid formed by the series of minor mergers at $1<z<2$. The galaxy therefore remains elliptical-like, with only a low-mass disk component, as in many real elliptical galaxies (see \citealp{Krajnovic2008}).

\medskip

As for the color evolution, the galaxy rapidly reddens after the $z>$1 mergers that converted it into an elliptical-like system (Figure~\ref{color}). During more than 4~Gyr, it maintains a red color with $B-V=0.7$--0.8 most of the time, which is consistent with colors of local ETGs according to \cite{Roberts1994}. The integrated color is indeed dominated by the contribution of the relatively old stars formed before/during the series of mergers. Because of the low star formation rate, young stars with bluer colors are not numerous enough to impact the integrated color, even though their contribution is maximized by neglecting dust extinction in our color estimates.

The same color evolution is also found in other color indices such as $U-B$ and $U-V$ (see Appendix C).

This elliptical galaxy remains constantly red during the quiescent phase in spite of a significant gas fraction left over after the initial mergers and constant infall of gas at a steady rate. Indeed, no mechanism removing the cold gas and/or terminating the accretion of fresh gas (like AGN feedback or virial shocks) occurs in our models. As a consequence, this red elliptical hosts a dense and massive disk of cold gas (Figure~\ref{galax_10Gyr}): in average $5\times10^9$~M$_{\sun}$ of cold gas with typical surface densities of 5--10~M$_{\sun}$~pc$^{-2}$.

This simulation shows that the effect of morphological quenching can be strong enough to turn an ETG red without requiring the removal of cold gas reservoirs and the termination of gas supply.

We will show in Section~4 how the low efficiency of star formation in this gas-rich red elliptical is connected to the presence of a stellar spheroid instead of a disk-dominated galaxy. We will also show that the gas disk is stable against star-forming instabilities, but unstable against non-axisymmetric modes which trigger the turbulence at the level required for $Q>1$ stability.

\subsubsection{Late blue phase}

The studied galaxy finally undergoes a major merger between $z=0.2$ and $z=0$, together with an increase of the diffuse gas accretion rate.

The combination of the major merger and the increased gas content allows a large starburst to take place, with a peak SFR of up to 35~M$_{\sun}$~yr$^{-1}$. As a consequence, the galaxy rapidly evolves toward blue colors.

However, this blue color at $z=0$ is only a transient phase related to the merger-driven starburst. We have let this galaxy evolve after $z=0$, and find that the star formation rate steadily decreases from 30~M$_{\sun}$~yr$^{-1}$ at $z=0$ to 10~M$_{\sun}$~yr$^{-1}$ after 2~Gyr. The galaxy reddens again, reaching $B-V>0.5$ within 2~Gyr (Figure~\ref{color}). This red early-type galaxy still contains a significant fraction of cold gas (6\% of the baryons), both because some gas is left over after the final merger, and some infalling gas that entered the virial radius at low redshift continues to feed the system (even if we neglect gas accretion at the virial radius after $z=0$). We thus find another ETG turning red spontaneously, under the effect of the MQ mechanism and without any mechanism removing its cold gas reservoir and/or stopping the infall of fresh gas.

To better understand the relation between morphology and color evolution, we have performed the same simulation without the final $z\simeq 0.2$ major merger -- earlier mergers and diffuse accretion remaining unchanged. While cosmologically realistic since directly motivated by a $\Lambda$-CDM simulation, the fiducial case may be less likely than this version without the final major merger, since the majority of massive galaxies do not undergo a major merger at $z<1$ \citep[e.g.,][]{genel08,Stewart2008}.

With this model, we find that it is the accumulation of cold gas accretion more than the final major merger itself that causes the return to blue colors at $z\simeq 0.2$ (Figure~\ref{color}). Indeed, even without the final major merger, the galaxy undergoes a phase of intense star formation at low redshift, with a star formation rate reaching 25~M$_{\sun}$~yr$^{-1}$ at $z=0$. The most important difference induced by removing the final merger lies in the morphology of the galaxy at $z=0$ (Figure~\ref{Evol}). With the final major merger, the galaxy is elliptical, and we find that its blue color is only transient and rapidly turns red under a MQ affect (see above). Without the final merger, the galaxy is more disk dominated, forming stars more durably; it develops spiral arms that host efficient star formation and are visible in the $z=0$ stellar disk (Figure~\ref{Evol}, see also Section~4). Hence, the gas disk that was stabilized against star formation during the quiescent red phase can become actively star-forming again if the accumulated gas mass increases too much.

\subsubsection{Summary}

Our main findings from the simulation so far can be summarized as follows.
First, a spheroid-dominated galaxy naturally becomes red and dead despite the
presence of a relatively massive gas disk. There is no need for an external
mechanism that removes the present gas or terminates the supply of fresh gas.
Second, a red spheroidal galaxy can turn back into a blue system composed of
a bulge and a disk. In our current simulation, this happened as a result
of a continuous growth of the disk over several Gyr through smooth gas
accretion. The rather smooth gas streaming into a galaxy could thus
significantly affect the morphology and the color of a galaxy. This is in
addition to the effects of mergers, which tend to build spheroids and thus
quench instability and star formation in gas disks.

\section{Disk stability}
\subsection{Stability analysis}
We find in our simulations that a dense gas disk can be stable against star formation when it is embedded in an early-type galaxy instead of a stellar disk. We will show in the following that a low gas surface density is not the reason for stability in the red ETG, and that the same disk would form stars efficiently if it was embedded in a stellar disk instead of a stellar spheroid.
\begin{deluxetable*}{ccccccccc}
\tabletypesize{\scriptsize}
\tablecaption{Comparison of galaxy properties at different instants in the simulation. The values corresponding to 13.7 Gyr are taken from the simulation without the final major merger, so that a gas disk can be identified.\label{table_disks}}
\tablewidth{0pt}
\tablehead{
\colhead{ }& \colhead{SFR (M$_\odot$ yr$^{-1}$)}& \colhead{$B-V$ }& \colhead{M$_g$ ($10^9$ M$_\odot$)}& \colhead{$\sigma_{\rm g}$ (km s$^{-1}$)}& \colhead{$\sigma_{\rm s}$(km s$^{-1}$)}& \colhead{$<\rm Q_{\rm g} >$} &\colhead{$<\rm Q_{\rm s} >$}& \colhead{$<\rm Q_{\rm tot} >$ }
}
\startdata
t = 3.5 Gyr& 9.1 & 0.39 & 3.96 & 25 & 43 & 8.7 & 4.6 & 2.3\\
t = 9 Gyr& 1.5 & 0.74 & 4.03& 16 & 26& 10.9 & 10.1 & 5.2\\
t = 10 Gyr &1.4 & 0.76& 6.14 & 15 & 19 & 9.1 & 7.6 & 4.2\\
t = 13.7 Gyr & 27.2 & 0.30 &11.23&22&33&8.2&2.3&2.2
\enddata
\end{deluxetable*}

\begin{figure}
\vskip 0.5cm
\centering
\includegraphics[width=8cm]{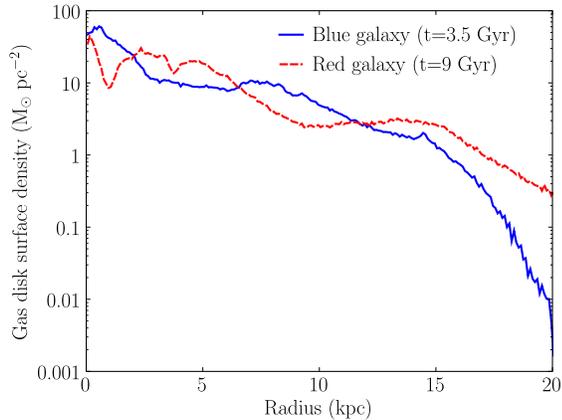}
\caption{Surface-density profiles of cold disk gas in the
early blue phase (solid, blue) and in the MQ phase (dashed, red).
The times are $t=3.5$ and 9~Gyr respectively.
The two disk profiles are similar, while the corresponding colors and SFRs are very different.
}\label{figdisks}
\end{figure}

The cold gas disks in the red elliptical galaxy at $t=9$~Gyr and in a blue spiral at $t=3.5$~Gyr have similar masses, radii and densities (Figure~\ref{figdisks}), but the former forms stars at a low rate, while the latter forms stars efficiently 
\footnote{If star formation is calibrated so that the spiral galaxy follows the Schmidt-Kennicutt relation, the galaxy during the MQ phase then lies below this relation, but within the observed scatter. Indeed, this scatter is rather high, and allows SFRs to differ by a factor of 10 or more for a given gas surface density. On the other hand, changing the SFR by such a factor is enough for a galaxy to turn red.}.
 The total (gas+stars) Toomre parameter for these two disks has an average value of 5.2 in the former case (red ETG), versus 2.3 in the latter one (blue disk galaxy) (Table \ref{table_disks}). The difference in the stability of these gas disks and their efficiency of star formation does not result from different gas masses or densities, but from a different depth of the gravitational potential measured by $\kappa$ \footnote{$\kappa$ is computed in radial bins of 100 pc from a smoothed version of the rotation curve.} and from a different contribution of stars $\Sigma_{\rm s}$ to the total disk density.

Note that the disk instability and the high level of star formation at $t=3.5$~Gyr are not artifacts due to relaxation of initial conditions. Indeed, the galaxy has been evolved in isolation for 500 Myr before being introduced in the simulation, which is a long enough time for the initial disk to acquire a realistic structure in the global galactic potential (see more details in \citealp{Martig2008}).
\smallskip

\begin{figure*}
\centering
\includegraphics[width=13cm]{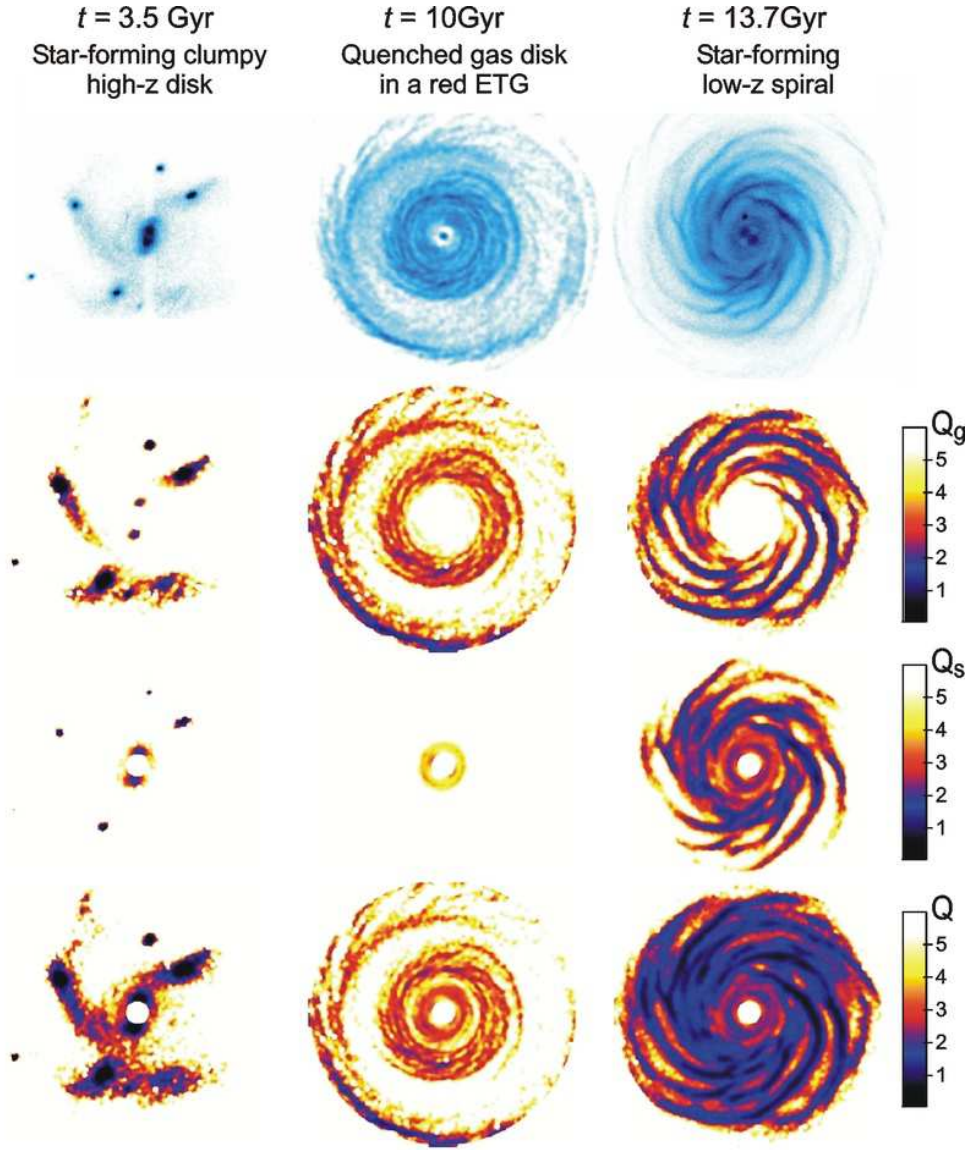}
\caption{Gas surface density and $Q$ parameter in the disk.
Shown are gas surface density (top), $Q_{\rm g}$, $Q_{\rm s}$ and the combined effective $Q$ in the three phases of evolution.
The box side is 30~kpc.
During the MQ phase, the effective $Q$ is above 2 across most of the disk, but below 3-4 in extended areas. This explains the stability of
axisymmetric modes and the lack of bound clumps in this phase, while
non-axisymmetric perturbations are clearly visible.
In the early and late phases, $Q$ drops below unity in several extended areas,
consistent with the appearance of strong axisymmetric perturbations
and bound clumps in these phases.
The effective $Q$ is similar to $Q_{\rm g}$ in the MQ phase, where
$Q_{\rm s}$ is rather high. In the unstable phases, especially the late one, the stellar disk component helps destabilizing the disk.
}\label{Qmap}
\end{figure*}

Maps of the Toomre parameter for the gas component ($Q_{\rm g}$), the stellar component ($Q_{\rm s}$) and the combined parameter $Q$ are shown in Figure~\ref{Qmap} for the blue galaxy at $t$=3.5~Gyr, for the inefficiently star-forming red ETG at $t$=10~Gyr (i.e. during the MQ phase), and the actively star-forming disk galaxy at $t$=13.7~Gyr in the case without the final merger so that a disk can be identified. These maps confirm that the star-forming disks have a low $Q$ parameter, with averaged values of $Q$ around 2 in the star-forming regions, and local overdensities with $Q<1$ in clumps and spiral arms. Regions with $Q\sim1$ are unstable against axisymmetric instabilities, reach high gas densities, and form stars actively.

\begin{figure*}
\centering
\includegraphics[width=12cm]{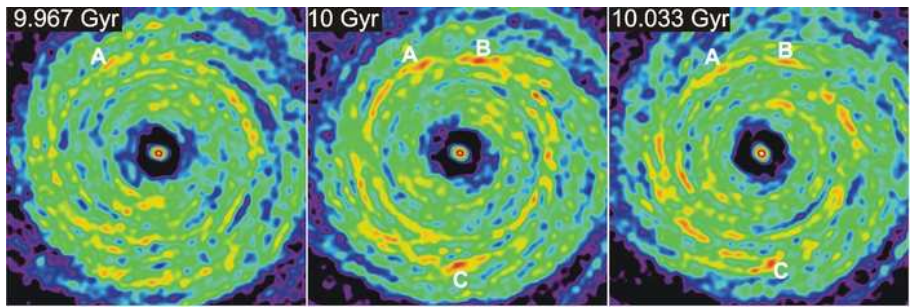}
\caption{Surface density maps of the gas disk during the MQ phase.
The snapshots are 33~Myr apart $t=10$~Gyr.
The box side is 15~kpc. The spiral armlets are strongly sheared and short-lived. For instance the dense structure A is just starting to form on the first snapshot, and already dissolving on the last one; structures B and C are not seen yet on the first snapshot, and already weaken on the last one.
}\label{Dmap}
\end{figure*}

The gas disk in the red ETG differs by having somewhat larger values of $Q_{\rm g}$ in its dense regions, because of the change in $\kappa$, and more importantly by having $Q_{\rm s}>>1$, because of the lack of stellar contribution $\Sigma_{\rm s}$ to the disk density and self-gravity. As a result, the combined $Q$ parameter remains significantly above 1 everywhere in the disk, preventing axisymmetric instabilities, even though non-axisymmetric armlets are present in the disk (Figure~\ref{Dmap}). During the MQ phase, the gas thus does not reach high densities, and the star formation efficiency is low: considering the global cold gas mass present in the galaxy, the SFR could have been expected to be ten times higher.

During the red phase of our simulation, the gas mass and surface density increase by a factor of 3. Yet the SFR remains constantly low and the galaxy remains red: the stability parameter $Q$ indeed remains large, and no dense gas clumps form. Only when the mass of the cold gas disk has increased above $9 \times 10^{9}$~M$_{\sun}$, the disk becomes unstable again, so that the star formation increases and the galaxy turns back blue: we find an average $Q \simeq 3$ with the densest regions reaching $Q=1$ at $t=11$~Gyr, when the galaxy begins to turn back to blue colors. This is illustrated by the simulation where the system evolves without the final major merger: the accreting cold gas continues to settle into the disk, increasing its surface density. Grand-design spiral arms develop (Figure~\ref{Evol}), and efficient star formation occurs in these arms. This results in a blue galaxy with a well-defined spiral structure. Here again we note that the star formation activity and the morphological evolution remain connected: when the galaxy turns back blue, it forms a massive stellar disk and rapidly takes a spiral disk+bulge morphology.

\subsection{Turbulence in the stable phase}

We have shown above that an early-type galaxy tends to stabilize its gas disk against star formation. However, this analysis holds only if the gas turbulence provides the necessary pressure support. If the velocity dispersion $\sigma$ was reduced in early-type galaxies, their gas disks could be Jeans-unstable (low $Q_{\rm g}$) in spite of the absence of a massive stellar disk. The velocity dispersion of the gas disk particles in our simulation remains about similar between the star-forming and non-star forming phases, where it lies on average around 15~km~s$^{-1}$ (Table \ref{table_disks}). For star-forming galaxies, Tamburro et al. (2009) indicate a characteristic HI linewidth value of 8-12~km~s$^{-1}$ in 11 spiral galaxies. This value includes turbulent velocity dispersion as well as thermal broadening, but turbulence is expected to dominate at least in the central (star-forming) regions of these galaxies. The velocity dispersion of our sticky particle medium is slightly higher\footnote{Note that particle noise and grid discretization make the velocity dispersion uneasy to measure in a differentially rotating disk, our 15~km~s$^{-1}$ value is thus an upper limit, but careful tests have ensured that the actual dispersion is not lower than 10~km~s$^{-1}$.} but includes all gas phases, so is consistent with observations on the blue sequence. A more important wonder is the level of turbulence in early-type galaxies with low star formation efficiencies.

\begin{figure*}
\centering
\includegraphics[width=12cm]{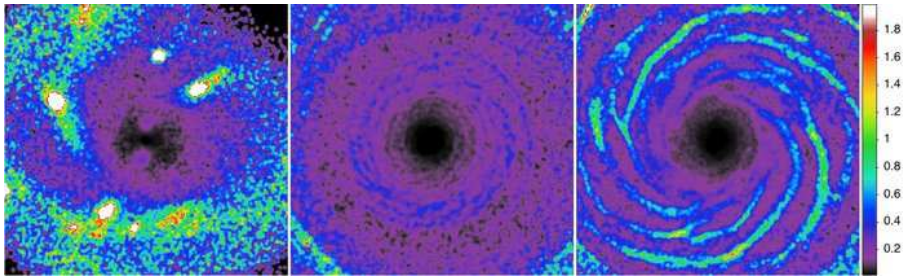}
\caption{
Ratio of gas density to the tidal density in the three phases
of evolution, at $t=3.5$, 9 and 13.7~Gyr respectively (for the simulation
without the final major merger). The box side is 30 kpc.
The early and late phases show extended regions where the ratio is above unity,
so self-gravity wins over tidal stripping and allows the growth of
perturbations.
The MQ phase shows no such regions, explaining the absence of bound clumps in this phase.
}\label{Tmap}
\end{figure*}
The ISM turbulence in star-forming spiral galaxies is mostly triggered by gravitational instabilities and star-formation feedback \citep{Elmegreen2003,wada,tasker,agertz}. An equally high level of turbulence in an inefficiently star-forming disk could then seem unexpected. Several processes can nevertheless maintain the turbulence in the absence of star-forming instabilities: these could be (i) gravitational instabilities without star formation, (ii) rotational shear powering turbulence \citep{shore03}, (iii) interactions with small satellites, dark subhalos, or more generally clumpy infall \citep{DSC09}, and (iv) the magnetorotational instability in magnetized disks \citep{Sellwood1999}. Mechanism (ii) does not occur in our simulations since the critical Reynolds number above which rotational shear instabilities might be able to grow is largely beyond reach \citep{Ji2006,Lesur2005}. Mechanism (iv) is not included in our models, and mechanism (iii) cannot dominate as the mass infall on the studied red galaxy is quite smooth. Mechanisms (i) should then be the main source of turbulence.

Gravity-driven turbulence does not necessarily require axisymmetric instabilities like dense clumps, and can actually be triggered with $Q>1$. A study of flocculent spirals by \citet{Elmegreen2003} shows that turbulent motions are generated by non-axisymmetric instabilities (spiral arms). This turbulence then acts at smaller scales where it form overdensities (clouds), which in turn collapse if they exceed the critical density for $Q$=1. In this picture, the turbulent energy is mostly pumped by large-scale, non axisymmetric instabilities like spiral arms. These structures can form with values of $Q$ much higher than 1 and trigger the turbulence, without involving the collapse of gas clumps where stars would form efficiently. Density maps of the non-star-forming gas disk at $t=10$~Gyr support this picture (Figure~\ref{Dmap}): small spiral arms are ubiquitous. They have small pitch angle typical for low-mass disks in shearing potentials, but none are circular rings. They resemble the very winded, flocculent structures observed in real ETGs, and could trigger turbulent motions just like in the star-forming spirals studied by \cite{Elmegreen2003}. But the Toomre parameter remains above 1 (around 2) in these arms (Figure~\ref{Qmap}), and the disk density remains below the tidal density (see below and Figure~\ref{Tmap}). No dense clump can then collapse and the star formation efficiency remains low. This is confirmed by the densest regions in these arms being short-lived and not collapsing into bound clumps, as shown by the evolution of the gas density over a short period (Figure~\ref{Dmap}).
\medskip

The velocity dispersions of our modeled gas disk in a red ETG are thus likely driven by one or several realistic processes including non-axisymmetric sheared instabilities, which does not seem to be a numerical artifact. The resulting turbulent pressure can provide the support against star-forming axisymmetric instabilities.
Observations of relatively high velocity dispersions in the non-star-forming outer HI disks of spiral galaxies \citep{Tamburro2009}, with level of velocity dispersions close to that of star-forming disks, suggest that some processes do drive the turbulence\footnote{The velocity dispersion in these outer HI disks could however be mostly thermal  \citep{Elmegreen1994}.} without triggering star formation, and could likely act in the internal regions of early-type galaxies to stabilize their gas disks in the morphological quenching mechanism.

\subsection{Star formation and resolution}

In the red ETG, the stability of the gas disk against axisymmetric perturbations prevents the formation of very dense gas clouds, for which high star formation efficiencies would result. We indeed assumed in our simulations that the star formation rate depends on the local gas density following a Schmidt law, which is realistic for various triggering mechanisms (pure gravitational instabilities, turbulent distribution of the density \citep{Elmegreen2002}, cloud-cloud collisions \citep{Smith1980}), and we impose a gas density threshold for star formation at space densities of 0.03~M$_{\sun}$pc$^{-3}$, which is about one atom per cubic centimeter. This is a conservative assumption: for a gas disk of height 200~pc, it corresponds to a surface density of 6~M$_{\sun}$pc$^{-2}$: this is the minimal density for cool diffuse atomic clouds formation, which is only the first step towards star formation. The critical density threshold for star formation is generally above this minimal density (see the review in \citealp{Elmegreen2002} and observations by \citealp{kennicutt89}).

We have performed a test simulation with a star formation threshold set to 0.25~M$_{\sun}$pc$^{-3}$. The star formation history and color evolution are globally unchanged (Figure~\ref{color}), and the colors are even slightly redder between 9~Gyr and 11~Gyr (with a maximum $B-V=0.90$ at $t=11$~Gyr). In the last part of the simulation, the galaxy experiences the same kind of bluing in both simulations. We also ran a simulation with no threshold for star formation, where the galaxy still becomes red during the quiescent phase (see Appendix A.4 for more details).

\medskip

That actively star-forming instabilities do not arise during the red early-type phase is likely a robust result, not only because $Q>1$, but also because the density of the gas disk remains below the limiting tidal density for the total galactic potential (Figure~\ref{Tmap}).
This means that even the densest regions of the gas disk, in the spiral armlets, cannot collapse because the tidal field exerted by the large-scale potential exceeds their internal gravity. The inefficient star-formation of the gas disk in the early-type galaxy is also consistent with this prescription, and the evolution of the gas density on short timescale (Figure~\ref{Dmap}) shows that no bound clouds form even in the densest parts of the spiral armlets.

The $\sim 250$ pc scale-height of the stable gas disk in our simulation may be an artifact of the $130$ pc spatial resolution.
Had the disk been thinner, the gas would have been denser and less stable against star formation according to the tidal density criterion. 
However, such a very thin disk would require a much lower level of velocity dispersion, which would result only in small, low-mass star forming regions

More generally, the mass resolution of a few $10^4 {\rm M}_\odot$ makes it impossible to resolve or rule out the formation of dense, 
star-forming gas clouds with sizes of a few tens of parsec and masses of a few $10^3 {\rm M}_\odot$.
The formation of small clouds in stabilized gas disks needs to be investigated in detail using simulations with a pc-like resolution, 
which cannot yet be achieved in cosmological zoom-in simulations. Nevertheless, we can infer that any such star-forming instabilities
would have been of small masses, sizes, turbulent speeds and star-formation rates compared to the classical star-forming regions of spiral 
galaxies.

\smallskip

The MQ mechanism unveiled in our simulations remained unnoticed in previous studies of galaxy formation. 
This led for instance \cite{khalatyan08} to conclude that AGN feedback was likely necessary to explain the reddening
of ETGs. According to the analysis above, an accurate description of the color evolution of galaxies
with gas disks relies on resolving star-forming clumps, which implies that the spatial resolution
should be much better than one kpc, and the mass resolution should be at least $\sim 10^5 {\rm M}_\odot$.
The turbulent velocity dispersion that acts to stabilize the disks should also be correctly simulated, which 
implies that shear instabilities such as spiral arms must be resolved. 
This requires the presence of a cold gas phase below $10^3$K.
While such a phase is naturally present in our sticky particle models (see Appendix B),
it requires a spatial resolution better than 100 pc in hydrodynamical simulations.
A low spatial or mass resolution or a high temperature floor in the ISM modeling might prevent the formation of
substructures and leave the gas disks smooth in all galaxies,
making it impossible to resolve dense star-forming clouds.

\section{Discussion}

\subsection{Morphological quenching in observed ETGs?}
\label{obs}

A fundamental difference between the MQ mechanism and the classical quenching mechanisms like AGN feedback and virial shocks is that red galaxies are not supposed to have their cold gas reservoirs removed or heated. A prediction if MQ plays an important role is that red early-type galaxies could contain significant amounts of cold gas, up to several percent of their baryonic mass. These cold gas reservoirs could sometimes reach masses and densities that would lead to efficient star formation and blue colors in spiral galaxies but do not form stars (or only inefficiently) when they are embedded in an elliptical galaxy.

In the Local Volume, the gas content of red ETGs (ellipticals and S0s) is often less than 1\% of the baryonic mass, if detected at all (e.g., \citealt{Bregman1992} for atomic gas, \citealt{Combes2007} for molecular gas). However, a large fraction of local ETGs are in the Virgo cluster or in dense groups, so they should have lost their cold atomic and molecular gas via ram pressure stripping \citep[but see][for the effect of ram pressure on molecular clouds]{tonessen}, harassment \citep{moore96}, strangulation \citep{kawata08}, etc.. Recent studies suggest that HI detections could actually be common among field ETGs \citep{morganti06}. The gas fractions remain generally low \citep{grossi09}, and a significant fraction is often in low-density outer structures or debris from tidal tails rather than internal disks \citep{oosterloo02,morganti06,oosterloo07}, but a variety of processes can be responsible for the low remaining gas masses (gas exhaustion, ram pressure and tidal stripping, strangulation) and the gas fraction could have been higher a few Gyrs ago.

Nevertheless, dense gas is sometimes found in the internal regions of ETGs, even in the form of molecular gas \citep[e.g.,][]{crocker08a, crocker08b,youngbl09, youngbl08}. Massive red ETGs can contain large amounts of dense gas in spite of their red colors, and only low levels of star formation are observed in these cases. A noticeable example is the elliptical galaxy NGC~2320 \citep{youngbl08}. This very red object ($B-V\simeq$~1) contains about $5 \times 10^9$~M$_{\odot}$ of molecular gas in a disk of a few kpc of radius, not counting any possible atomic gas component, but its star formation rate remains at undetected levels, while such a high mass of molecular gas would turn any spiral galaxy blue. ESO~381-47 is another red ETG that has an unusually low star formation efficiency in its gas disk, as reported by \cite{donovan09}. \cite{morganti06} show that NGC~3414 and 4278 host large HI disks but only old stellar populations (see also Shapiro et al. in preparation).

Although they do not dominate in the Local Universe, these examples of early-type galaxies with massive cold gas disks but red integrated colors and low star formation activities are likely candidates for the MQ mechanism.

\medskip

If the morphological quenching mechanism plays a statistically important role in forming the Red Sequence, a prediction is that red early-type galaxies would frequently host significant amounts of cold gas, of few percent of their baryonic mass. This would not necessarily be the case in the Local Universe where other environmental processes have exhausted the gas content. This could be more frequent at high ($z>1$) or intermediate ($z\sim0.5$) redshifts, when most ETGs are not in virialized clusters and dense groups yet, and the cold gas infall rates are higher. Recent stellar population studies of red galaxies at high-redshift \citep{M09,T09} indicate that their colors are consistent with some residual star-formation activity. This could be compatible with the presence of gas disks forming stars inefficiently in a MQ mechanism, instead of complete removal of cold gas and termination of gas supply. \citet{kriek09} have recently obtained a near-infrared spectrum of one of the red ETG in their $z\sim$~2 sample, which suggest a low but non-zero star formation rate. This could also be consistent with the presence of a morphologically-quenched gas disk. This red ETG has a stellar mass of 2$\times10^{11}$~M$_{\sun}$ and an estimated SFR of 1--3~M$_{\sun}$~yr$^{-1}$, which compares favorably to the MQ phase in our simulations, but at higher $z$.

\subsection{Observed blue ellipticals}

At the end of our simulation, we obtain a blue elliptical galaxy just after a major merger. This is also briefly the case at $t\simeq6$ Gyr. These are galaxies that have just been transformed into ellipticals by mergers and experienced a starburst at the same time. A galaxy that has undergone a starburst can indeed remain blue for one or two Gyrs depending on the strength of the starburst and the photometric bands considered (e.g., Larson et al. 1980).

Such blue elliptical galaxies are rare in the Local Universe. Lee et al. (2008) find that 4\% of ETGs at $z<0.1$ have blue colors, and most of these blue ETGs are low mass, dwarf ellipticals. There are nevertheless known cases of red ETGs with little on-going star formation but that have undergone a starburst in the last billion year: these E+A galaxies are rare at $z=0$, probably because galaxies that have undergone a starburst-triggering merger in the last couple of Gyrs are rare \citep[e.g.,][]{genel08}. The blue elliptical at the end of our simulation results in fact from the combination of two different phenomena, a major merger at $z=0.15$, and a high gas accretion rate between $z=0.4$ and $z=0$ so that the final merger involves a gas-rich galaxy. Our simulation shows that this can happen in $\Lambda$-CDM cosmology, although the rapidly decreasing major merger rate and gas accretion rate at $z<1$ \citep[e.g.,][]{Keres2009} should make such cases infrequent, in agreement with the low fraction of E+A galaxies at low redshift. At higher redshift, the higher merger rate could increase the fraction of blue ellipticals and E+A galaxies, resembling for instance the blue elliptical at $z\sim 1.5$ following several minor mergers.

The two blue ETGs in our simulations are short-lived: their recently formed stellar spheroid is quenching their star forming activity, and they are due to become red ellipticals in about 1~Gyr. When we continue the simulation after the $z=0$ instant, the final galaxy also turns back red.

\subsection{Revived blue disks?}

We found in our model a case of red spheroidal galaxy evolving into a blue galaxy with a massive disk. In the model without the final $z\simeq 0.2$ major merger, this results in a disk+bulge galaxy at $z=0$ with a grand-design spiral structure. However, our model does not support a "spiral rebuilding scenario" according to which the majority spiral disks would be re-built after major mergers from tidal material falling back and/or fresh infalling gas \citep{Hammer2005}, because:

(1) The reformation of a star-forming disk requires a high gas infall rate ($\simeq 10$~M$_{\sun}$~yr$^{-1}$) during several Gyr, without any significant mergers during this period. This is above the average accretion rate expected from $z\sim 1$ to $z\sim0$ for a typical galaxy with the mass of the Milky-Way \citep{Keres2009}, and above the typical observed gas infall rates, too \citep{Sancisi2008}. Most galaxies would then be expected to accrete lower amounts of gas: in this case, the MQ mechanism remains active and the star formation is quenched, preventing the re-formation of a stellar disk and the return to blue colors. A "spiral rebuilding" mechanism below $z\sim 1$ then requires high rates of diffuse gas infall that seem unlikely to take place in most galaxies; the return of a red galaxy to blue colors is a possible mechanism but likely taking place only in a small fraction of galaxies.

(2) The re-built disk galaxy keeps important signs of the high-redshift mergers, even in our model without the final major merger. A massive, red stellar halo, as large as $\sim$20 kpc in radius, is present, amounting to about 25\% of the stellar mass at $z=0$. The kpc-sized bulge amounts to 35\% of the stellar mass, only 40\% of the stellar mass hence being in the rotating disk. This disk galaxy rebuilt after a series of mergers therefore differs from typical local spiral galaxies \citep{Binney1987}, that have a significantly lower mass fraction in the rotating disk component.

Furthermore, the transient blue elliptical phases in our models, either at $z \simeq 1.5$ or $z \simeq 0$, correspond to post-merger elliptical galaxies that are about to redden rapidly. Hence, neither of the blue ETGs found in our $\Lambda$CDM-based models have blue colors persisting more than 1~Gyr. They are not signs of spiral galaxies being rebuilt after the mergers (but see \citealp{Kannappan2009}), and these blue elliptical phases, potentially E+A-like, result from the time needed for a stellar population to redden after the star formation is quenched.

\section{Conclusion}

We proposed a straightforward mechanism for the quenching of star
formation in galactic disks once they become dominated by a stellar spheroid, which stabilizes the disk against fragmentation into bound, star-forming clumps. We demonstrated the natural occurrence of such morphological quenching in a cosmological simulation.

For the purpose of a numerical study, we used a new efficient technique for simulating the evolution of a galaxy in a cosmological context. The merger history and smooth accretion history are recorded for a given galaxy in a cosmological simulation, and the evolution of the target galaxy is re-simulated at high resolution from $z=2$ to $z=0$, including gas dynamics and star formation.

We have studied the morphological and photometric evolution of a field galaxy ending with a stellar mass of 2$\times$10$^{11}$~M$_{\odot}$ at $z=0$. This galaxy goes through three different phases of growth: (a) a blue phase from $z\simeq 2$ to $1$, during which a stellar spheroid is built by a series of minor mergers, (b) a long red phase from $z\simeq 1$ to $0.2$, with smooth gas accretion at
a high rate $\simeq 10$~M$_{\sun}$~yr$^{-1}$, and (c) a blue phase after $z \simeq 0.2$, following a major merger. Based on the analysis of the evolution of this galaxy, our main conclusions are as follows.

Once a galaxy acquires a spheroid-dominated morphology,
by a major merger, minor mergers or disk instabilities,
it can spontaneously turn red. This occurs without gas removal or suppression of gas supply by other quenching mechanisms such as AGN feedback, virial shock heating, gravitational quenching or ram pressure stripping.

This morphological quenching in early-type galaxies reflects the stabilization of the disk by the dominant presence of a pressure-supported stellar spheroid, replacing the stellar disk. The stabilization of the gas disk is a result of two effects: the steep potential well induced by the spheroid and the removal of the stellar disk component that used to enhance the self-gravity of perturbations in the disk.
Early-type galaxies can thus be red and dead even in the presence of a gas disk, either left over after mergers or accreted later from the cosmological reservoir. The MQ mechanism should be equally efficient if massive stellar spheroids are formed by another process than mergers, for instance giant disk instabilities and clump coalescence at high redshift \citep{noguchi99,EBE08,DSC09,Ceverino2009}.

In the case of merging galaxies with a higher gas fraction than what we studied here, the final disk to spheroid ratio of the remnant is increased \citep{Hopkins2009}, so that the effect of MQ would be weaker. However, most of these remnants are bulge-dominated, so that we can still expect the bulge to have a stabilizing effect, unless the disk is particularly massive. Generally speaking, we propose that a galaxy with most of its stars in bulge and/or a thick disk might experience some morphological quenching, which could not only apply to ellipticals but also to S0 galaxies. Future simulations will probably allow us to study this in more details.

We also find that a red-and-dead galaxy can turn back into a blue star-forming galaxy. The gas disk can fragment to clumps and form stars again if it becomes dense enough to overcome the stabilizing effect of the spheroid. This requires a long phase of smooth gas accretion at a high rate, which is possible even at $z <1$ but only for a small fraction of the galaxies. Such a galaxy that turns blue again is expected to have a spheroid of mass comparable to the disk mass, so it is not a likely progenitor of Milky Way-like spirals. Our models indicate that blue ellipticals and E+A galaxies are more likely to be recently formed spheroids that are about to turn red than red merger remnants that are about to re-create a massive blue disk.

The natural process of morphological quenching provides a simple
explanation for the most distinct feature of the observed galaxy bimodality \citep{DB06}, namely the fact that the color evolution of a galaxy is tightly linked to its morphological evolution.
The MQ can help quenching the star formation in galaxies on the Red Sequence. It should be naturally working in concert with other quenching mechanisms, depending on the mass and environment of each galaxy. But a unique feature of MQ is the fact that it can explain the presence of red-and-dead galaxies in halos below the critical mass for virial shock heating, $\sim 10^{12}{\rm M}_\odot$, where the other quenching mechanisms fail. 

Current observations of large atomic and molecular gas reservoirs in local early-type galaxies, with red colors and low star formation rates,
suggest that the reddening of elliptical galaxies cannot be solely a result of gas removal or suppression of gas supply. Morphological quenching could thus contribute to the formation of the modern Red Sequence. The presence of quenched gas disks in red ETGs may be more frequent at higher redshift, where gas fractions and gas accretion rates are typically higher.

\section*{acknowledgments}
We thank the anonymous referee for constructive comments and suggestions.
We are indebted to Bruce Elmegreen for useful discussions on gravity-driven turbulence and star formation, and to Alison Crocker for useful discussions on molecular gas and star formation in early-type galaxies. We thank Damien Le~Borgne for help with stellar population modeling, and Martin Bureau, Fran\c{c}oise Combes, Eric Emsellem, S\'{e}bastien Fromang, Ryan Joung, Kristen Shapiro, and Lisa Young for useful comments and discussions. We are grateful to Thomas Guillet for his help with Python.
Simulations were carried out at CEA/CCRT computing center.
This work has been supported at CEA by the Horizon project, a
France-Israel Teamwork grant, and by the Agence Nationale de la Recherche under contract ANR-08-BLAN-0274-01. It has been partially supported at HU by grants from ISF, GIF, DIP, and the HU Einstein Center and at UCSC by NASA ATP.


\appendix 

\section{Appendix A. Simulation of galaxies in their cosmological context}\label{appendix_sim} 

In the following subsections, we describe the initial cosmological simulation, the extraction of merger and accretion histories, and the technique used to perform the high-resolution re-simulation and to analyze it.

\subsection{A.1. Cosmological simulation}
The cosmological simulation has been performed with the Adaptive Mesh Refinement code RAMSES \citep{Teyssier2002}. The simulation box contains 512$^3$ dark matter particles and has a comoving length of 20~h$^{-1}$~Mpc. This gives a mass resolution of 6.9$\times$10$^6$~M$_{\sun}$, so that a Milky Way type halo is made of a few 10$^{5}$ particles. The cosmology is set to $\Lambda$-CDM with $\Omega_m$=0.3, $\Omega_{\Lambda}$=0.7, H$_0$=70~km~s$^{-1}$~Mpc$^{-1}$ and $\sigma_8$=0.9.

\subsubsection{A.1.1. Halo finding}
The dark matter halos are detected in the cosmological simulation using the HOP algorithm \citep{Eisenstein1998}, with $\delta_{\rm peak}$=240, $\delta_{\rm saddle}$=200 and $\delta_{\rm outer}$=80. The minimal number of particles per halo is fixed to 10 (which corresponds to a minimal halo mass of 6.9$\times$10$^7$~M$_{\sun}$).

Particles that do not belong to a gravitationally bound halo will also be taken into account; we will refer to them as ``diffuse'' mass accretion when they are accreted by the studied halo.

\subsubsection{A.1.2. Selection of the studied halo/galaxy}
We call ``main halo'' the halo (and galaxy) which we want to study and re-simulate, chosen in the final $z$ = 0 snapshot. When the studied halo is located in a rather empty environment (i.e. it does not belong to a cluster, and no more massive halo is found within 2~Mpc at $z=0$), we do not take into account the large-scale tidal field when we re-simulate the history of the halo. This is the case for the halo studied in the present paper (see right panel of Figure \ref{FigCosmo}).

\subsubsection{A.1.3. Merger and accretion histories extraction}
The method used to extract the merger and accretion histories consists in following the main halo (more exactly the most massive progenitor of the final halo) from $z = 2$ to $z = 0$ and recording all the halos -- the {\em merger history} -- and diffuse particles -- the {\em accretion history} -- crossing a fixed spherical boundary around it. This boundary is a sphere of radius equal to the final virial radius of the main halo at $z=0$, which is defined as:
$$
R_{\rm vir}= \left(\frac{3 M_{\rm tot}}{4\pi\times 200 \rho_c}\right)^{\frac{1}{3}}
$$
where $M_{\rm tot}$ is the mass of the main halo at $z=0$.

At $z=2$, the radius of this sphere is larger than the initial main halo: it encloses not only this main halo but also some satellite halos and diffuse particles. The re-simulation will begin at $z=2$ with all the halos and diffuse particles initially enclosed in the boundary. To this aim, we record their mass, position, velocity, and spin for the satellites, all in the rest frame of the main halo. Positions are expressed in physical units instead of comoving, but with velocities corrected for the Hubble flow: a radial velocity $v_r=H(z)\times d$ (where $d$ is the distance to the center of mass of the main halo) is added to the velocity of diffuse particles and to the velocity of the center of mass of the halos.

We then check each subsequent output of the cosmological simulation (snapshots are 60 Myr apart), and search for particles that are inside the boundary but have not been found there earlier-on: these are particles that have crossed the boundary since the last snapshot. After having undergone the same modifications as described previously (i.e. coordinate transformation from comoving to physical and addition of the Hubble flow), their mass, position and velocity are written in a catalog (together with the redshift of their entrance in the boundary).

As far as halos are concerned, we record a halo if at least one particle belonging to it has crossed the boundary, and write in the catalog the position, velocity, mass and spin of this halo when its first particle(s) have crossed the boundary. The center of mass of the halo is thus still outside the boundary, particularly in the case of a large halo, but this choice has the advantage of being straightforward and accurate. According to its orbital parameters, the halo then might not totally enter the boundary and merge with the main halo. This is not a problem for the re-simulation, where it will not merge with the main halo either and will simply make a fly-by.

The last point is that we have to handle the case of particles that enter a companion halo while this one has already been written in the catalog (i.e. the halo has grown since some of its particles have first crossed the boundary and this halo still has not merged with the main halo). In this case, the particles are handled as diffuse accretion. A halo can then be surrounded by a cloud of diffuse particles. This approximation concerns only the "satellites of a satellite" merging simultaneously, which concerns only a small fraction of the total infalling mass.

To summarize, the final products of this analysis are the following:
\begin{itemize}
 \item an initial conditions file: a list of all the halos and particles that are already inside the boundary (the final virial radius) at $z=2$
 \item a catalog: a list of all the particles that cross the boundary between $z=2$ and $z=0$ and of all the interacting companion halos.
\end{itemize}

\subsection{A.2. High-resolution re-simulation}
We now replace each halo of the cosmological simulation by a galaxy made up of gas, star and dark matter particles, and each diffuse particle of the cosmological simulation with a blob of lower-mass, higher-resolution gas and dark matter particles. They start interacting with the main galaxy following the orbital and spin parameters given by the cosmological simulation.

The code we use for the re-simulation is the Particle-Mesh code described in Bournaud \& Combes (2002, 2003). The density is computed thanks to a Cloud-In-Cell interpolation and the Poisson equation is then solved using Fast Fourier Transform techniques. Time integration is made using a leapfrog algorithm with a time step of 1.5~Myr. The maximal spatial resolution (and softening length) is 130~pc. The mass resolution is 1.4$\times$10$^5$~M$_{\sun}$ for stellar particles initially present in the galaxies, 2.1$\times$10$^4$~M$_{\sun}$ for gas particles and stellar particles formed during the simulation, and 4.4$\times$10$^5$~M$_{\sun}$ for dark matter particles.

Gas dynamics is modeled with a sticky particle scheme with $\beta_r$=0.8 and $\beta_t$=0.7:these are the factors by which the relative radial and tangential velocities of colliding gas particles are reduced after the collision. Star formation is computed with a Schmidt-Kennicutt law \citep{kennicutt98}: the star formation rate is proportional to the gas density to the exponent 1.5, above a fixed threshold of 0.03~M$_{\sun}$pc$^{-3}$, which is about one atom per cubic centimeter. The efficiency of star formation is chosen so that at $z=2$ the main galaxy would consume its gas with a timescale of 1.5~Gyr, consistent with the star-formation timescale of high-redshift galaxies \citep{Daddi2007}.

Feedback from supernovae explosions is not included. Indeed, the Schmidt-Kennicutt law, which is an observed relation, should already account for this effect. What is more, simulations have shown that feedback only has a minor role in quenching star formation (Kravtsov 2003), except maybe in the case of very strong starbursts where strong winds could be triggered. We have also chosen not to include feedback from AGNs. AGN feedback could be implemented in our re-simulation technique just as in cosmological zoom simulations, but in the present paper we present a quenching mechanism that can take place independently of AGN activity.

\subsubsection{A.2.1. Model galaxies}

Each halo recorded in the dark matter-only cosmological simulation is replaced with a realistic galaxy (Figure 10), made of a disk (gas and stars), a stellar bulge and a dark matter halo \footnote{Note that no hot gas halo is included}. The disks have a Toomre profile, the bulge is a Plummer sphere and the halo follows a Burkert profile \citep{Burkert1995,Salucci2000}. This halo profile has been chosen because ``real'' galaxies most probably have halos with a core rather than with a cusp. The fact that dark matter halos have a flat core is widely observed in the Local Universe \citep{Borriello2001,Salucci2001,deBlock2003,Oh2008,Trachternach2008} and high redshift observations \citep{FS06} suggest that the mass in the inner regions of galaxies at z$\sim$2 is essentially baryonic (which might not be the case if dark matter halos had a steep profile).

The core radii of our dark matter halos are chosen according to the scaling relations proposed by Salucci \& Burkert (2000) and the halos are truncated at their virial radius.

\begin{figure}
\centering
\includegraphics[width=6cm]{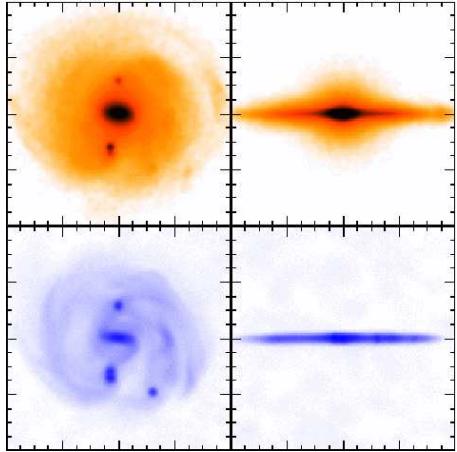}
\caption{Initial distribution of stars (top panel) and gas (bottom panel) for the main galaxy, seen face-on and edge-on (each panel is 40~kpc~x~40~kpc in size). The small overdensities that can be seen belong to the disk, and are the result of gravitational instabilities due to the high initial gas fraction.}\label{InitialG1} 
\end{figure}

\begin{figure}
\centering
\includegraphics[width=7cm]{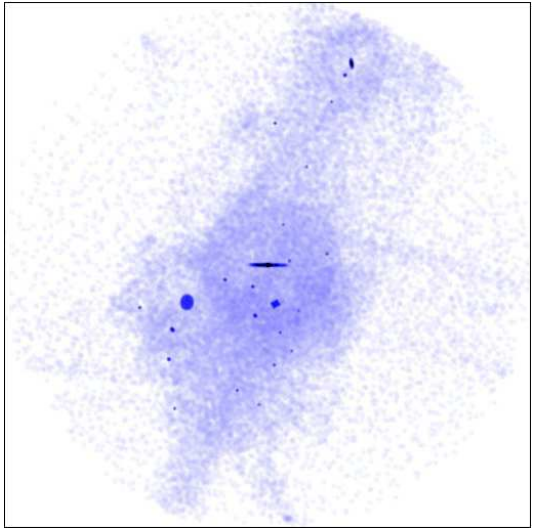}
\caption{Large scale view of the gas distribution at the beginning of the simulation (the panel is 440~kpc~x~440~kpc in size). The disk of the main galaxy is seen edge-on and is surrounded by satellite galaxies and diffuse gas as prescribed by the cosmological simulation at $z=2$. }\label{InitGas} 
\end{figure}

\begin{deluxetable}{ccccccccccc}
\tabletypesize{\scriptsize}
\tablecaption{Characteristics of model galaxies.\label{table_galax}}
\tablewidth{0pt}
\tablehead{
\colhead{ } & \colhead{G1}& \colhead{G2} & \colhead{G3} & \colhead{G4} & \colhead{G5} & \colhead{G6}& \colhead{G7} & \colhead{G8} & \colhead{G9}& \colhead{G10} 
}
\startdata
\cutinhead{Dark matter halo}
Mass (M$_{\sun}$)& 1.7$\times10^{11}$& 5$\times10^{8}$&1$\times10^{9}$ & 2.5$\times10^{9}$&5$\times10^{9}$& 7.5$\times10^{9}$& 1$\times10^{10}$ &2.5$\times10^{10}$ & 7.5$\times10^{10}$ & 4.1$\times10^{11}$\\
N$_{\rm particles}$ & 400000& 1136& 2272& 5681&11363 & 17045& 22727 & 56818& 170454& 931818\\
R$_{\rm trunc}$ (kpc)& 38.63& 8.21& 10.34 & 14.03& 17.68& 20.24& 22.27& 30.23& 43.60& 102.41\\
R$_{\rm core}$ (kpc)& 25.80& 1.10& 1.40& 2.20& 3.00& 3.60& 4.10& 6.40& 10.80& 20.70\\
\cutinhead{Stellar disk}
Mass (M$_{\sun}$)& 2.5$\times10^{10}$& 8$\times10^{7}$&  1.5$\times10^{8}$ & 3.8$\times10^{8}$& 7.7$\times10^{8}$& 1.1$\times10^{9}$& 1.5$\times10^{9}$ & 3.8$\times10^{9}$ & 9.18$\times10^{9}$ & 5.5$\times10^{10}$\\
N$_{\rm particles}$ & 170078& 523& 1047&2618 & 5236&7855 & 10473&26183 & 62615& 376915\\
R$_{\rm trunc}$ (kpc)& 6.00& 0.32& 0.45& 0.71& 1.01& 1.24& 1.43& 2.26& 3.91& 9.15\\
R$_{\rm carac}$ (kpc)& 3.00& 0.16& 0.22 & 0.36& 0.50& 0.62& 0.71& 1.13& 1.96& 4.57\\
h (kpc)& 0.17& 0.01& 0.01& 0.02& 0.03& 0.03& 0.04& 0.06& 0.11& 0.26\\
\cutinhead{Gas disk}
Mass (M$_{\sun}$)& 7.5$\times10^{9}$& 2.3$\times10^{7}$&4.6$\times10^{7}$ & 1.1$\times10^{8}$&2.3$\times10^{8}$& 3.5$\times10^{8}$& 4.6$\times10^{8}$& 1.1$\times10^{9}$ & $2.8\times10^{9}$ & 8.3$\times10^{9}$\\
N$_{\rm particles}$ & 357165& 1099& 2199& 5498& 10997& 16495&21994 & 54985& 131493& 395760\\
R$_{\rm trunc}$ (kpc)& 9.85& 0.52& 0.74& 1.17& 1.66& 2.03& 2.35& 3.71& 6.43& 15.03\\
R$_{\rm carac}$ (kpc)& 8.13& 0.43 & 0.61& 0.97& 1.37& 1.68& 1.94& 3.06& 5.31& 12.42\\
h (kpc)& 0.07& 0.01& 0.01& 0.01& 0.01& 0.01& 0.02& 0.03& 0.05& 0.12\\
gas fraction& 0.30& 0.30&0.30 & 0.30& 0.30& 0.30& 0.30& 0.30& 0.30& 0.15\\
\cutinhead{Stellar bulge}
Mass (M$_{\sun}$)& 2.6$\times10^{9}$& 0& 0 & 0&1.4$\times10^{6}$& 2.1$\times10^{6}$& 2.8$\times10^{6}$ &7.3$\times10^{6}$ & 2.9$\times10^{9}$ & 1.7$\times10^{10}$\\
N$_{\rm particles}$ & 18897& 0& 0& 0& 10& 15& 20&  52& 20871& 125638\\
R$_{\rm trunc}$ (kpc)&1.97 & 0& 0& 0& 0.33& 0.40& 0.47& 0.74& 1.28& 3.00\\
R$_{\rm carac}$ (kpc)& 1.11& 0&0 &0 & 0.19& 0.23& 0.26& 0.42& 0.73& 1.70\\
\enddata
\end{deluxetable}

The total mass of the galaxy is divided in 17\% of baryons and 83\% of dark matter (the dark matter mass is the mass of the halo found in the cosmological simulation). The gas content in the disk varies according to the redshift at which the galaxy is introduced in the simulation and according to its total mass: for small galaxies ($\rm M_{\rm halo}<10^{11}\rm M_{\sun}$) the gas disk represents 30\% of the stellar disk \citep{Roberts1994}, whatever the redshift, and for massive galaxies it is 30\% at ``high'' redshift ($\rm z > 0.8$) and 15\% at ``low'' redshift (this is in order to match the very high gas fractions observed at $z\sim2$ for instance by \citealp{Daddi2008} and the low redshift values of \citealp{Roberts1994}).

To decrease the computing time while keeping a realistic cosmological context, a different galaxy is not created for each halo of the cosmological simulation, but instead a set of model galaxies spanning the whole range of masses needed (and closely sampling the masses of the halos) is created (Table \ref{table_galax} shows the characteristics of these model galaxies). When a halo has to be introduced in the simulation, the model galaxy that is the closest in mass is chosen among this set (note that the most massive companions are exactly reproduced).

Each model galaxy is evolved in isolation for 500~Myr before being introduced in the simulation (star formation is shut down during this period so that the initial gas fraction is conserved). This allows the galaxies to acquire a realistic structure (e.g. spiral arms, bars...) before beginning to interact with one another, so that we can be sure that all the processes observed during the interactions are due to the interactions themselves and not to the growth of non-axisymmetric structures from axisymmetric initial conditions (which is not what we are interested in).

\subsubsection{A.2.2. Diffuse mass accretion}

The catalog we have extracted from the cosmological simulation also takes into account diffuse accretion of dark matter particles onto the main halo. In the re-simulation, we replace each of these diffuse particles by a blob of gas particles (17\% in mass) and dark matter particles (83\% in mass). Each blob is thus made of 65 gas particles and 15 dark matter particles.

These particles are randomly distributed in a sphere. The radius of this sphere is chosen such that the mean density of the blob is 100$\rho_c$, which is chosen to be lower than the local density of the dark matter halo of the main galaxy so that the blob can be tidally disrupted when approaching the main galaxy: this allows the diffuse gas density field to be smooth, at least in its the densest regions (Figure~\ref{InitGas}). To respect this density criterion, the blobs introduced in the simulation before $z=1$ have a radius of 2.5~kpc, the blobs introduced between $z=1$ and $z=0.5$ have a radius of 3.3~kpc and the blobs introduced after $z=0.5$ have a radius of 5~kpc.

The particles belonging to the blob are given no velocity dispersion since we expect the blob to be disrupted by the tidal field of the main galaxy anyway.

\subsubsection{A.2.3. Carrying out the simulation}

We start the simulation with the main galaxy (galaxy G1 in Table \ref{table_galax}) at $z=2$ and place around it the satellite galaxies and gas blobs as prescribed by the initial conditions file (see Figure \ref{InitGas} the result of this operation). To place a satellite galaxy, we choose among the model galaxies the closest in mass, and put it at the right location with the right velocity. Its orientation is set assuming that the direction of the angular momentum of the disk is the same as the direction of the angular momentum of the dark matter halo detected in the cosmological simulation.

Then, we read the catalog line per line and add in the simulation the other galaxies and the other blobs at the exact moment when they appeared in the cosmological simulation, with the exact position, velocity and spin extracted from the cosmological simulation.

\subsection{A.3. Analysis}

\subsubsection{A.3.1. Global mass evolution}

To study the evolution of the total mass of stars and gas in the main galaxy, we take into account all the particles located within a sphere of radius 25~kpc around the center of the galaxy. In the same way, to study the star formation rate of this galaxy, we only count stars that form within this sphere (i.e. we do not take into account stars forming in approaching companions that have not merged with the main galaxy yet).

\subsubsection{A.3.2. Stellar ages and colors}

Our simulation contains two types of stellar particles: those that were formed during the simulation (and for which determining an age is straightforward) and those that were initially present in the model galaxies we introduced in the simulation.  We need to give an age to these stars in order to be able to compute galaxy colors, and for that we have used and tested different assumptions. Their common characteristic is that they assume that the first stars began to form 500~Myr after the Big Bang (that is around $z=10$). The subsequent star formation histories have been tested (they concern not only the main galaxy, but also the infalling satellites for which we call $T_{in}$ the moment when they are introduced in the simulation):
\begin{itemize}
 \item an exponentially decreasing SFR with a timescale $T_{sf}$ given by $T_{sf}=T_{in} - 500$~Myr
\item an exponentially decreasing SFR with a timescale $T_{sf} = 1$~Gyr
\item a constant SFR between 500~Myr and $T_{in}$
\item half of the stars formed at $t = 500 $~Myr and the other half formed according to a constant SFR between 500~Myr and $T_{in}$
\end{itemize}

Knowing the age of a stellar particle, we use the spectral evolution model {\sc Pegase.2} to compute stellar colors assuming a Salpeter initial mass function \citep{salpeter} from 0.1 to 120~M$_{\sun}$ \citep{Fioc1999}. 
Each stellar particle is thus given a luminosity in B, V and K bands (Johnsson filters), all magnitudes being then computed in the Vega magnitude system.

We assume a uniform solar metallicity in all our computations. Indeed, massive galaxies at $z\sim 2$ are known to have a relatively high metallicity already, even if somewhat sub-solar in their outer disks \citep{Erb2006,Bournaud2008}, and the $B-V$ color (on which we are going to focus) is not significantly sensitive to the metallicity (contrary to some other bands, e.g. \citealp{Larson1980}). 

The integrated galaxy luminosity in various bands is then computed by summing up the stellar population spectra (we then consider all stars included in a sphere of radius 25~kpc around the center of this galaxy). Extinction is not taken into account, since the evolution of galaxies between the blue and red populations is known not to be caused by dust absorption effects; only some specific cases have their integrated color significantly reddened by extinction \citep{Alam2002,Unterborn2008}. Neglecting dust reddening is a conservative assumption regarding our main conclusion on the spontaneous reddening of elliptical galaxies. 

\subsection{A.4. Sensitivity to star formation threshold}

In the fiducial simulation shown in this paper, star formation is modeled with a Schmidt-Kennicutt law \citep{kennicutt98} above a threshold of 0.03~M$_{\sun}$pc$^{-3}$. This is a conservative assumption since this threshold corresponds to the minimal density for cool diffuse atomic clouds formation (see also Section 4.3). 
To test the influence of this choice on our results, the same simulation has been run with a higher threshold (0.25~M$_{\sun}$pc$^{-3}$, which is possibly more realistic) and with a threshold set to 0, which is unrealistic but useful for comparisons.

\begin{figure}
\centering
\includegraphics[width=8cm]{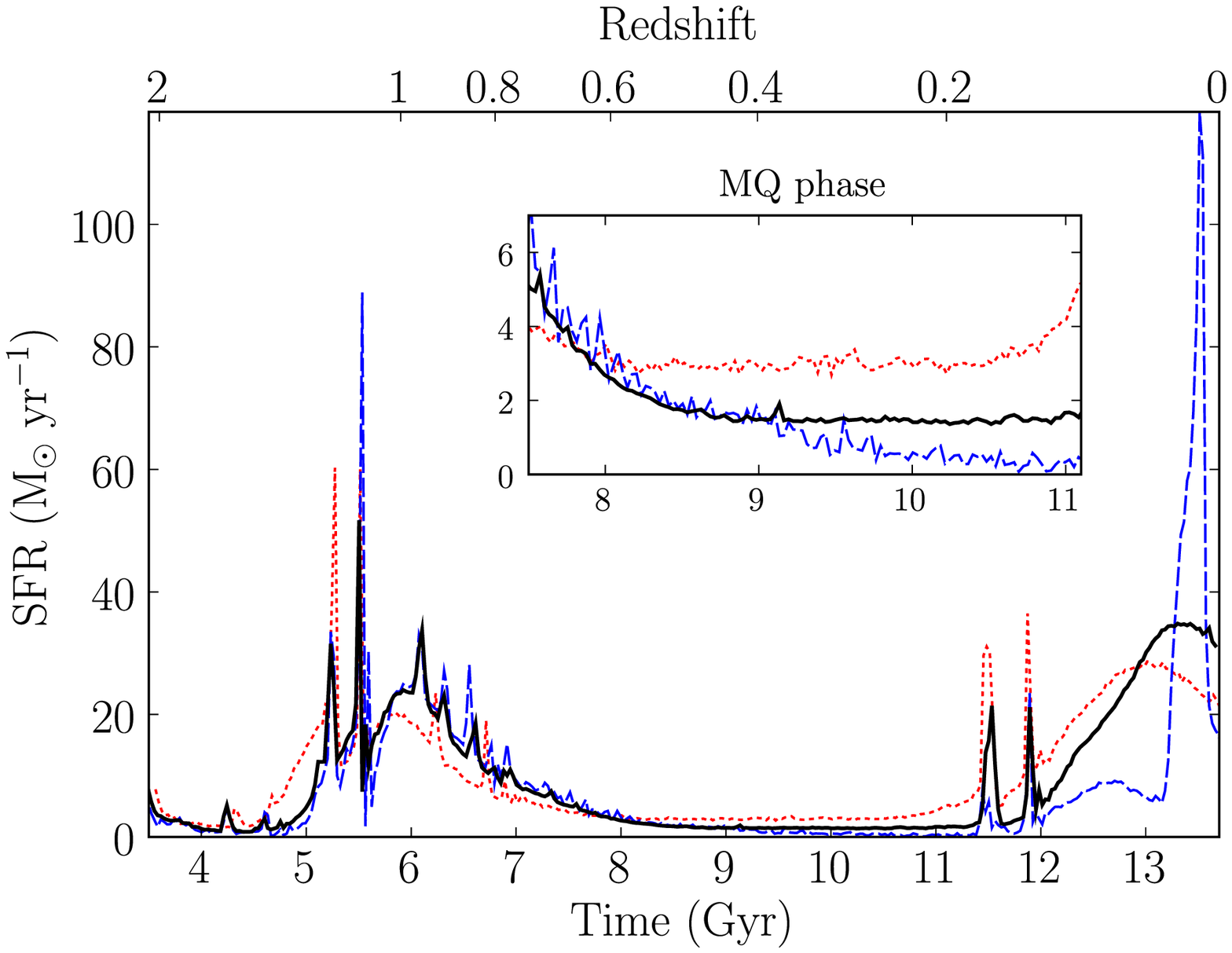}
\includegraphics[width=8cm]{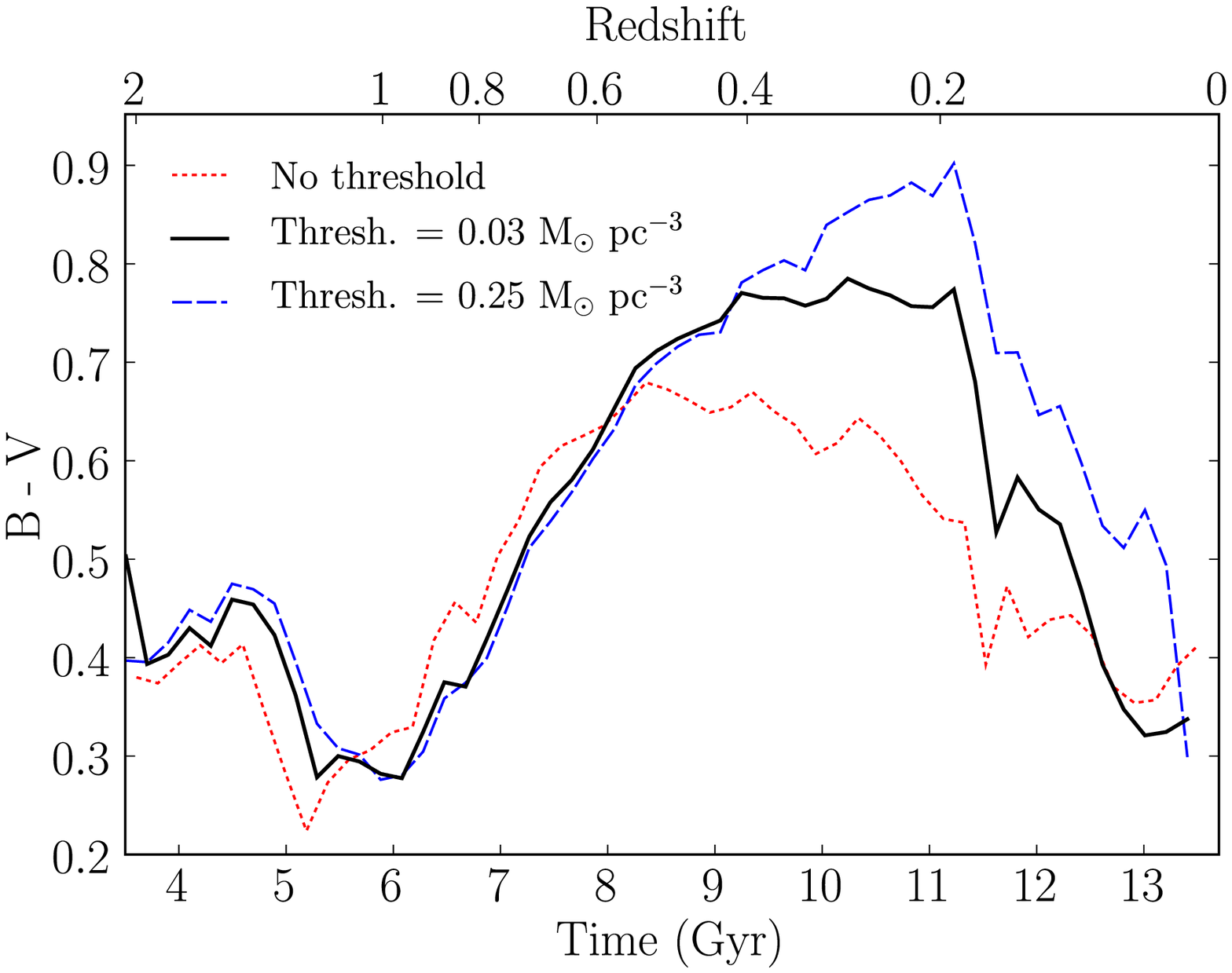}
\caption{Star formation rate (left panel) and corresponding color evolution (right panel) for three different star formation recipes : without any  threshold, with a threshold set to 0.03 M$_{\odot}$pc$^{-3}$ (our standard conservative assumption) and a threshold set to 0.25 M$_{\odot}$pc$^{-3}$ (a possibly more realistic assumption). During the morphological quenching phase, all models exhibit a lower star formation rate and a redder color, even if the effect is stronger for higher thresholds. }\label{comp_threshold}
\end{figure}

We find an overall similar evolution of the star formation rate in all cases (see left panel of Figure \ref{comp_threshold}). Most of the differences arise during the morphological quenching phase, where, when no threshold is used, the star formation rate is twice higher with respect to what was found for the fiducial simulation. Similarly, setting a higher threshold reduces the star formation rate during this phase. As a consequence, the main galaxy experiences a reddening during the morphological quenching phase for the three different models, but this reddening is stronger for higher star formation thresholds (see right panel of Figure \ref{comp_threshold}).

Our results on morphological quenching (that were obtained using a very conservative assumption on the value of the  star formation threshold) are thus robust. Using a higher (possibly more realistic) threshold would only increase the MQ effect.

\subsection{A.5. Sensitivity to initial conditions}

To test whether or not our results depended on the initial conditions chosen (i.e. for instance gas fraction or Hubble type of the main galaxy G1), we have run the exact same simulation replacing G1 with an elliptical galaxy. This galaxy is modelled as a Plummer sphere of stars (with a characteristic radius equal to 2~kpc and truncated at 6~kpc), with the same total baryonic mass as G1 (the dark matter halo is the same as for G1). This allows us to test the influence of the gas fraction and of the Hubble type at the same time.

In the first 1.5~Gyr of the simulation, we find a significant difference between this new simulation and the one presented so far. Indeed, since the initial galaxy now does not contain gas at the beginning of the simulation (Figure 13), its star formation rate is much lower and thus its color is much redder (see Figure \ref{col_ell}). 

\begin{figure}
\centering
\includegraphics[width=8cm]{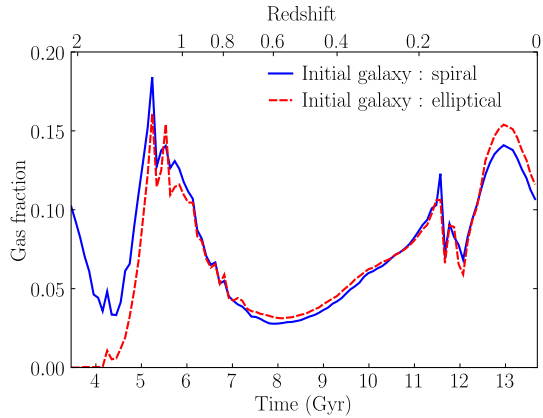}
\caption{Evolution of the dense ($\rho > 0.01$~M$_{\odot}$pc$^{-3}$) gas fraction (with respect to total baryonic mass) in a sphere of radius 25~kpc around the main galaxy, in the case where this main galaxy is a spiral at $z=2$ and in a case where it is an elliptical.}\label{fgas_ell}
\end{figure}

\begin{figure}
\centering
\includegraphics[width=8cm]{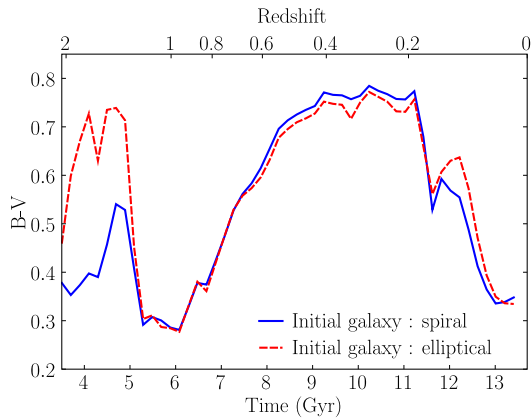}
\caption{Color evolution of the main galaxy in the case where this main galaxy is a spiral at $z=2$ and in a case where it is an elliptical: after 1.5~billion years, the color evolution is independent of the initial conditions.}\label{col_ell}
\end{figure}

Yet, the galaxy acquires gas very quickly (mainly thanks to accretion from the filaments) and after 1.5~Gyr the gas fraction becomes remarkably similar in both simulations (see Figure \ref{fgas_ell}). Following this, the SFR evolution is the same, and thus the color evolution is also the same (see Figure \ref{col_ell}). 

In fact, the change in the initial conditions only affects the very beginning of the simulation both in terms of morphology and color.
Our results are robust versus changes in the initial characteristics of the main galaxy since our mechanism to produce a red galaxy only requires:
\begin{itemize}
 \item a stellar spheroid, that is formed independently of the initial conditions (a disk will be destroyed by the first series of mergers, an elliptical will just stay elliptical and grow)
\item a gas disk that is here fueled by the filament and does not depend on the initial conditions either.
\end{itemize}

\section{Appendix B. The sticky particle scheme for gas disks dynamics}

Most of the gas mass in galactic disks is in a cold atomic phase and molecular clouds rather than in warm and hot diffuse phases \citep[e.g.,][]{brinks90}. This medium is very heterogeneous and its turbulent pressure dominates the thermal pressure \citep[see reviews in][]{Elmegreen2002,Burkert2006}. It cannot then be described as an ideal, continuous gas, and a description based on separate, interacting clouds is more adequate \citep{larson81}. A particle-based hydrodynamical code, like SPH codes, could offer such a description only if dense and cold small-scale substructures are resolved together with their turbulent motions. This cannot be achieved with the 130~pc softening length of our simulations\footnote{at fixed CPU time, an SPH simulation would actually have a lower spatial resolution and number of particles than our sticky particle simulations.}. A temperature floor of a few $10^3$ or $10^4$~K would have to be imposed, as would also be the case for grid hydrodynamic codes (AMR) if a spatial resolution of a few parsecs cannot be achieved. The ISM modelled with such codes would then remain dominated by a high thermal pressure with a minimal sound speed of $\sim 10$~km~s$^{-1}$, and homogeneous on scales of at least 1~kpc. Such a description would be inadequate for star-forming gas disks, and would be even more problematic regarding the stabilization of a gas disk in an ETG, as the minimal sound speed imposed would artificially stabilize such disks independently of their real turbulent support.

This motivates the use, in our galaxy-scaled re-simulations, of the sticky particle (SP) scheme, which is a realistic description for a collisional medium \citep{brahic77} or a cloudy, heterogeneous gas medium \citep{LR81}. In our SP code, gas particles interact with each other through inelastic particle-particle collisions. The velocity dispersion of particles represents the turbulent speed of the modelled gas medium - not its thermal sound speed. The cross-section given to gas particles ensures about one collision per particle and per vertical crossing time and corresponds to a filling factor in spiral disks of about 5\%, which can be realistic for the ISM \citep{dAB04}. Colliding particles bounce back, which models the turbulent pressure acting on gas substructures. These collisions are inelastic. We use a restitution factor $\beta = 0.75$ for velocities (about 50\% for the energy), because two gas clouds with similar masses and densities, surrounded by vacuum, that collide with each other would keep half of their initial kinetic energy in a kinetic form. The dissipated kinetic energy is converted into internal energy for each gas cloud (particle) and radiated away. With the collision rate of particles, this ensure that the velocity dispersion dissipates in about a vertical crossing time (in the absence of triggering source), as found in high-resolution hydrodynamical simulations of turbulence \citep{ML99}. The complete equations of the SP code can be found in \citet{BC02} and references therein.

The SP scheme is not intended at being adequate on large cosmological scales. However, it could also describe the hot gas phase if collisions between particles are considered as the analogs of collisions between atoms: SP could then also reproduce thermal pressure. It is nevertheless most frequently employed to model the dynamics of the cool and cold ISM in galactic disks \citep[e.g.,][]{schwarz81, noguchi99}. An extensive study of cloud formation and disruption in SP simulations has been performed by \cite{CG85}. Comparison of SP and SPH for disk dynamics can be found in \citet*{BCS05}. Parametric studies in \citet[][appendix A]{BC02}  and \citet{BCS05} show that over a certain range, variations of the dissipation parameter $\beta$ have limited effects on the global evolution of modeled gas disks as a stronger dissipation is compensated by a stronger gravitational heating. Motivations for using an SP scheme rather than SPH at the scale of galaxies, in particle-based simulations, are further reviewed in \cite{SC02}, who show that the formation and evolution of disk galaxies in the cosmological context can be realistically studied with an SP code.

\section{Appendix C. $U-B$ and $U-V$ color evolution}

\begin{figure}
\centering
\includegraphics[width=8cm]{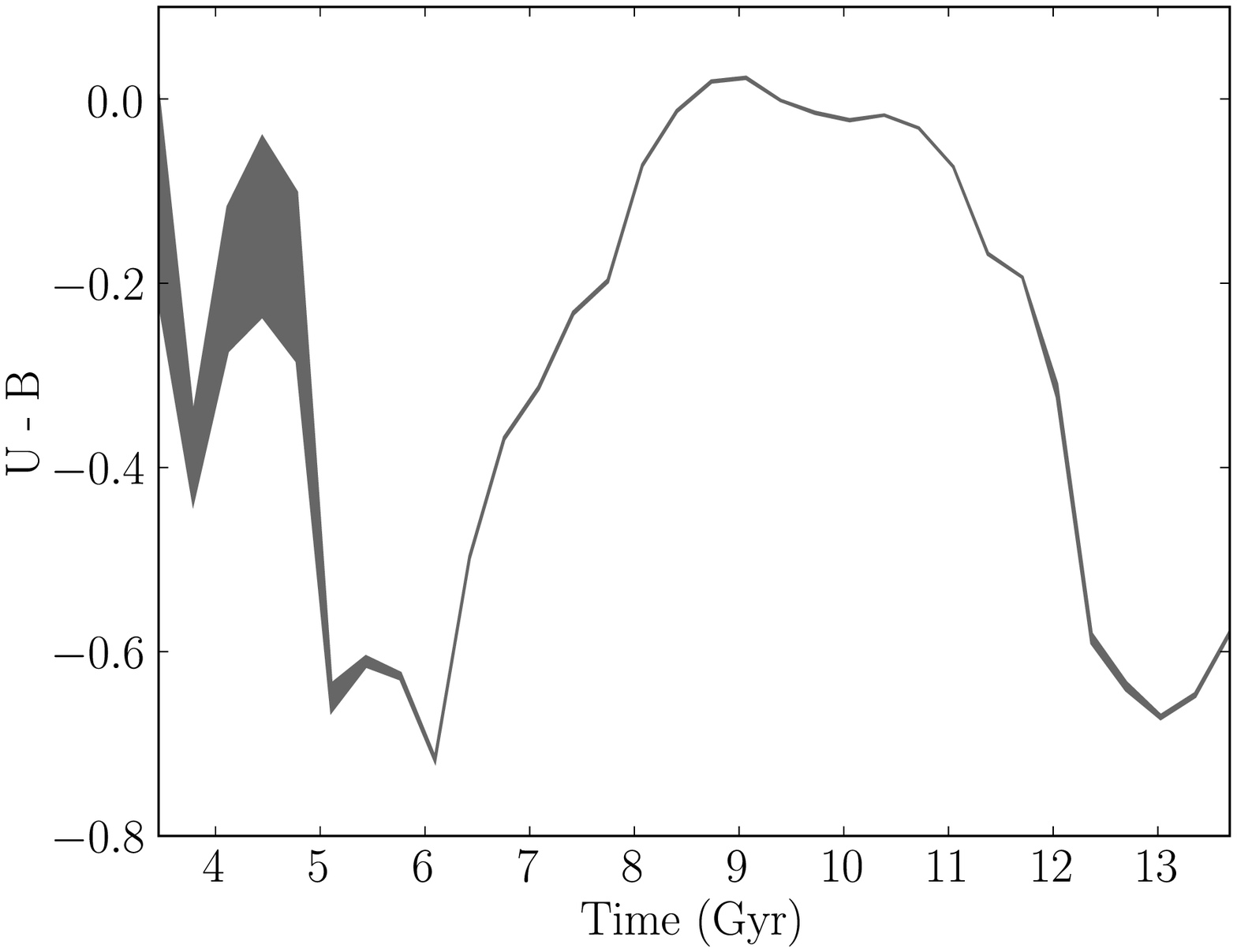}
\includegraphics[width=8cm]{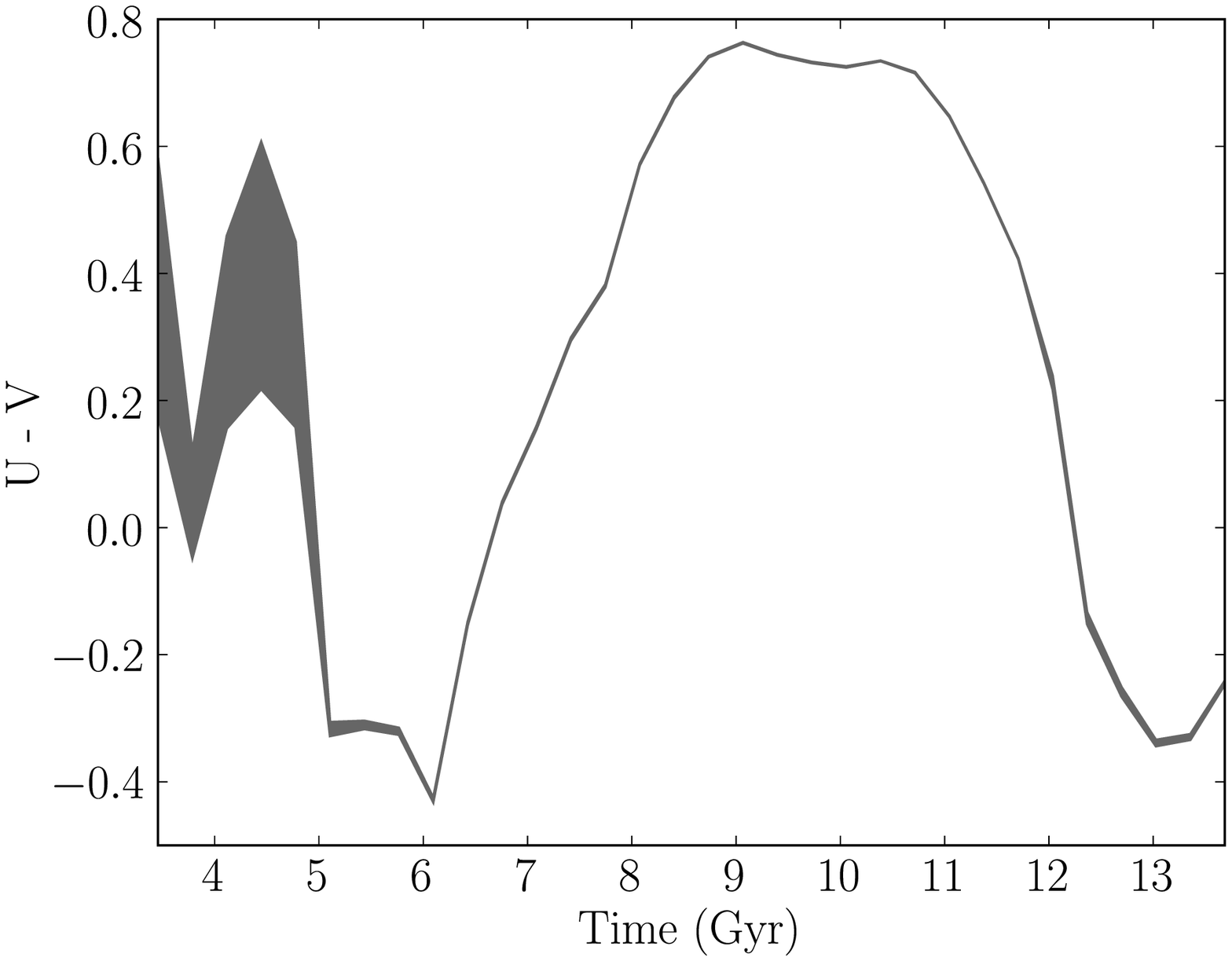}
\caption{U-B and U-V  color evolution. The shaded areas along the curves represent the range of colors 
obtained for different star formation histories (Appendix \ref{appendix_sim}). Both colors evolve similarly with time and in both cases a clear reddening is found during the morphological quenching phase. }\label{col_ub_uv}
\end{figure}

We have studied the $U-B$ and $U-V$ color evolution (see Figure \ref{col_ub_uv}). We find results that are in good agreement with what we deduced from the $B-V$ color evolution, in particular the reddening during the morphological quenching phase is a robust fact. For instance, $U-B$ grows from $-0.7$ during the initial starburst phase to 0 during the morphological quenching phase.

$U-B=0$ is not a very red color for a galaxy (for instance, \citealp{Weiner2005} define the red sequence as  $U-B>0.1$), but taking extinction into account could slightly increase this value and make it compatible with observations of red sequence galaxies, in particuler at high redshift.

\end{document}